\documentclass[aps,prx,reprint,superscriptaddress,nofootinbib]{revtex4-2}

\usepackage{amsmath,graphicx,color}
\usepackage{bm}
\usepackage{amsfonts}
\usepackage{amssymb}
\usepackage{bbold}
\usepackage[colorlinks=true,linkcolor=blue,citecolor=blue,urlcolor=blue]{hyperref}
\usepackage[all]{hypcap}

\newcommand{\gin}[0]{{\gamma_\mathrm{in}}}
\newcommand{\gex}[0]{{\gamma_\mathrm{ex}}}
\newcommand{\gpin}[0]{{\gamma'_\mathrm{in}}}
\newcommand{\gpex}[0]{{\gamma'_\mathrm{ex}}}

\renewcommand{\>}{\rangle}

\newcommand{\vect}[1]{\bm{#1}}
\usepackage{verbatim} 
\newcommand{\rev}{\textcolor{black}}

\newcommand{\mc}{\mathcal}

\begin{document}

\title{Topology protects chiral edge currents in stochastic systems}

\author{Evelyn Tang$^a$}
\email{evelyn.tang@ds.mpg.de}
\affiliation{Max Planck Institute for Dynamics and Self-Organization (MPIDS), D-37077 G\"ottingen, Germany}

\author{Jaime Agudo-Canalejo$^a$}
\email{jaime.agudo@ds.mpg.de}
\affiliation{Max Planck Institute for Dynamics and Self-Organization (MPIDS), D-37077 G\"ottingen, Germany}

\author{Ramin Golestanian}
\affiliation{Max Planck Institute for Dynamics and Self-Organization (MPIDS), D-37077 G\"ottingen, Germany}
\affiliation{Rudolf Peierls Centre for Theoretical Physics, University of Oxford, Oxford OX1 3PU, United Kingdom}

\date{\today}

\begin{abstract}
Constructing systems that exhibit time-scales much longer than those of the underlying components, as well as emergent dynamical and collective behavior, is a key goal in fields such as synthetic biology and materials self-assembly. Inspiration often comes from living systems, in which robust global behavior prevails despite the stochasticity of the underlying processes. Here, we present two-dimensional stochastic networks that consist of minimal motifs representing out-of-equilibrium cycles at the molecular scale and support chiral edge currents in configuration space. These currents arise in the topological phase due to the bulk-boundary correspondence and dominate the system dynamics in the steady-state, further proving robust to defects or blockages. We demonstrate the topological properties of these networks and their uniquely non-Hermitian features such as exceptional points and vorticity, while characterizing the edge state localization. As these emergent edge currents are associated to macroscopic timescales and length scales, simply tuning a small number of parameters enables varied dynamical phenomena including a global clock, dynamical growth and shrinkage, and synchronization. Our construction provides a novel topological formalism for stochastic systems and fresh insights into non-Hermitian physics, paving the way for the prediction of robust dynamical states in new classical and quantum platforms.
\end{abstract}

\maketitle

\def\thefootnote{a}\footnotetext{These authors contributed equally to this work.}

 \section{Introduction}

Why are biological functions carried out so robustly, even when the underlying components are stochastic in time and randomly distributed in space?  Living systems can have stable properties that endure for time-scales much longer than the lifetime of the underlying constituents, that contribute to memory and adaptive processes \cite{winfree1980geometry,Prost2015}. The emergence of stable and reproducible timescales and length scales is also a key goal in the design of synthetic biological systems \cite{2017144,schwille2018maxsynbio} or in the engineering of reconfigurable materials e.g.~through dissipative self-assembly \cite{deng2020,riess2020}. However, these strongly out-of-equilibrium systems often lack a comprehensive theoretical framework, which prevents us from understanding or describing these processes \cite{phystoday}.

\begin{figure*}
 \includegraphics[width=1\linewidth]{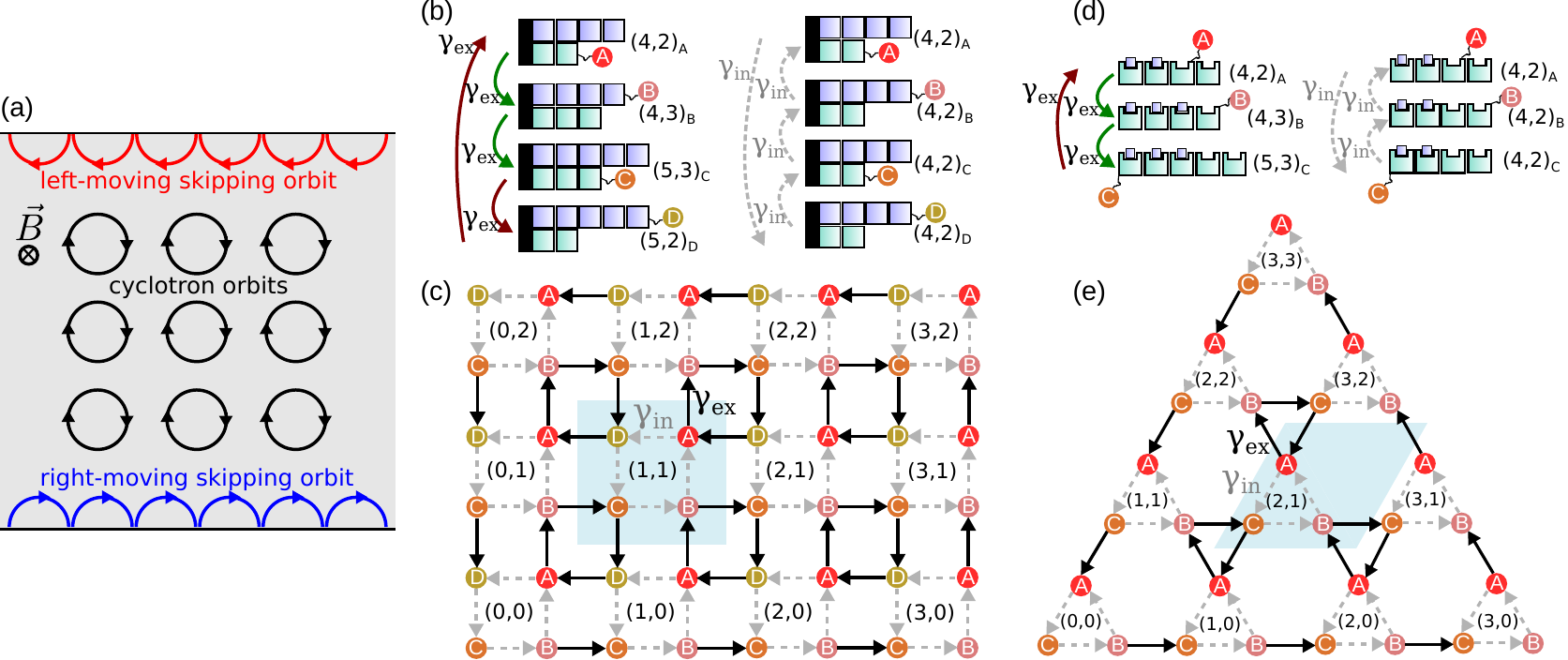}
 \caption{Minimal motifs and resulting lattices. (a) In the semiclassical picture of the quantum Hall effect, cyclotron orbits of electrons in the bulk, with chirality defined by the perpendicular magnetic field $\vec{B}$, result in skipping orbits at the edges, giving oppositely-directed currents at the top and bottom edge of the sample. (b) 4-state model for a structure composed of two types of monomers X and Y (violet and green). The configuration of the system is determined by the number of monomers $(x,y)$ and the internal state (A, B, C, or D), and changes of internal state are represented as tagging a given subunit and thus priming it for addition or removal. External transitions (governed by $\gex$) involve the addition (green arrows) or removal (red arrows) of subunits, whereas internal transitions (grey, $\gin$) cause relaxation of the internal state.  (c) Square lattice corresponding to the 4-state model. (d) 3-state model, describing a system in which subunits X (green) are modified e.g.~via phosphorylation (violet). The external transition from C to A involves the removal (red arrow) of a modified subunit. (e) Kagome lattice corresponding to the 3-state model. Note the similarity between the lattices (c,e) and the quantum Hall effect in (a), with clockwise external cycles (black, $\gex$) representing cyclotron orbits, and internal transitions (grey, $\gin$) enabling diffusive translation of these orbits. The skipping orbits in (a) correspond to the counter-clockwise trajectory along the system boundary in (c,e). Shaded blue square and rhombus in  (c) and  (e) respectively correspond to a unit cell in each lattice. \label{fig1}}
 \end{figure*}

As a step towards answering this question, we present stochastic models that exhibit emergent dynamical phenomena reminiscent of many observations prevalent in biology \cite{vanZon7420,Chang2012,Zhang2020,Mitchison1984,Leibler1993}, such as a global clock, dynamical growth and shrinkage, as well as synchronization. In our models, all these phenomena hinge on the emergence of a topologically-protected chiral edge state in the stochastic system. To illustrate this connection, we use the canonical example of a topological phase: the quantum Hall effect \cite{Cohen2019}, see Fig.~\ref{fig1}(a). Here, electrons make cyclotron orbits in the bulk that correspond to unidirectional edge states at the boundaries. The system transport is hence dominated by the propagating edge states which are exponentially localized at the boundaries and protected from disorder and perturbations. The robustness of these edge currents make them a desirable feature for the support of stable emergent  phenomena. However, they have yet to be realized in biochemical systems, i.e.~systems governed by memoryless (Markovian) classical stochastic dynamics.  While topological states showing stationary polarization have been recently reported in one-dimensional stochastic systems \cite{muru17,DasbiswasE9031}, states with propagating edge currents have not yet been reported. 

 The two-dimensional networks we introduce are constructed from simple repetitive motifs, which correspond to out-of-equilibrium cycles at the molecular scale, and form the analog of cyclotron orbits in the quantum Hall system. In biochemical networks, such microscopic transitions are common due to out of equilibrium transitions that consume a fuel such as ATP or GTP. Many of these appear to leave the system unchanged while consuming energy, and having been dubbed ``futile cycles'' are ubiquitous in biology \cite{Hopfield1974,Samoilov2005}.  Beyond the analogy to quantum Hall physics, the chiral edge currents in our system imply motion along the boundaries of configuration space rather than real space. As such, they enable oscillations (e.g.~cyclical conformational changes of a protein complex, or assembly and disassembly of a biopolymer) governed by physical constraints in the system rather than the specific timescales of the underlying microscopic transitions, which do not need to be fine-tuned \cite{winfree1980geometry,phystoday, Prost2015}. Hence, these systems constitute an excellent example of the ``structure determines function'' paradigm of biology. Besides their possible relevance to biological oscillators, our models could guide the engineering of synthetic nonequilibrium machines that perform work (e.g.~stochastic low Reynolds number swimmers \cite{RG2008,RG2010}).

The out-of-equilibrium or nonreciprocal stochastic transitions render the transition matrix non-Hermitian. While much of the formalism to describe topological states of matter was initially developed for quantum electronic systems \cite{PhysRevLett.42.1698,PhysRevLett.62.2747,PhysRevLett.120.146402,PhysRevLett.116.133903,Xiong_2018,PhysRevX.9.041015,PhysRevB.97.045106,PhysRevB.100.075437} and later extended to classical systems such as mechanics \cite{Kane2014,Serra-Garcia2018,PhysRevLett.122.118001}, photonics \cite{PhysRevX.4.021017,Benalcazar61,doi:10.1021/acsphotonics.8b00117,PhysRevA.100.032102}, acoustics \cite{PhysRevLett.119.255901,Ni_2017}, electrical circuits \cite{Imhof2018,PhysRevLett.122.247702}, active matter \cite{PhysRevX.7.031039,shankar2020topological,DasbiswasE9031}, and population dynamics \cite{2020arXiv200901780K,yoshida2020chiral}, this formalism has traditionally been reserved for Hermitian systems that respect energy conservation and isolation from the environment. As our systems are strongly dissipative and break detailed balance at the microscale, we build a topological description that is explicitly non-Hermitian. Non-Hermitian physics is an area of great recent interest \cite{PhysRevLett.120.146402,PhysRevA.100.032102,Heiss:2012dx,PhysRevLett.116.133903,Xiong_2018,PhysRevLett.123.205502,You19767,PhysRevX.10.041009,PhysRevB.97.045106,PhysRevE.93.042310,PhysRevX.9.041015,2020arXiv200601837A,doi:10.1021/acsphotonics.8b00117}, and can demonstrate unique properties that have no analog in Hermitian systems. \rev{A celebrated example is that of exceptional points: singularities where eigenstates coalesce \cite{Heiss:2012dx,PhysRevLett.116.133903,Xiong_2018,PhysRevLett.123.205502,You19767,PhysRevX.10.041009}, in contrast to degenerate level crossings in Hermitian systems where eigenvectors remain distinct. A second example is vorticity, a topological winding of eigenvalues in complex space \cite{PhysRevLett.120.146402}, differing from topological invariants in Hermitian systems which are typically defined using eigenvectors, as in these systems the eigenvalues are purely real.}
We show that our models support the topological Zak phase \cite{PhysRevLett.62.2747,Benalcazar61,doi:10.1021/acsphotonics.8b00117,PhysRevX.4.021017,PhysRevLett.119.255901,Imhof2018,PhysRevB.100.075437}, but only support propagating edge currents in the non-Hermitian case. Further, they exhibit the above-mentioned non-Hermitian features such as exceptional points and vorticity, creating a novel theoretical formalism for stochastic systems distinct from previous proposals that do not show these features \cite{DasbiswasE9031,muru17}.

The paper is organized as follows. In Sec.~\ref{sec:emergence}, we describe the minimal motifs that can be used to construct two-dimensional stochastic networks with emergent chiral edge currents (Sec.~\ref{sec:minimal}), and show how global cycles arise using stochastic simulations and analytical calculations (Sec.~\ref{sec:edgecurrents}). In Sec.~\ref{sec:topology}, we generalize the model to explore the transition from Hermiticity to non-Hermiticity, and explain how the emergence of edge states can be understood as a topological transition. This is a consequence of the Zak phase (Sec.~\ref{sec:berry}) and uniquely non-Hermitian topological invariants such as vorticity (Sec.~\ref{sec:nonhermitian}), which results in exponential localization at the edges (Sec.~\ref{sec:localization}). Finite-size effects and the effects of disorder on topological protection are discussed in Sec.~\ref{sec:realistic}. In Sec.~\ref{sec:complex}, we apply our model to specific examples in soft and living matter, such as biomolecular oscillators (Sec.~\ref{sec:globalclock}) and stochastic swimmers (Sec.~\ref{sec:swimmer}), and further extend the model to account for other biologically-relevant features, such as asymmetry in configuration space (Sec.~\ref{sec:asymmetric}) and shared boundaries between different subsystems (Sec.~\ref{sec:coupled}), which lead to new behaviors such as dynamical instability and synchronization. Finally, in Sec.~\ref{sec:comparison} we put our work in the context of previous proposals for topological protection in related systems.  We end with a discussion of the implications and future directions for our work.

\section{Emergence of chiral edge currents \label{sec:emergence}}

\subsection{Minimal motifs \label{sec:minimal}}

We consider discrete stochastic processes that operate in a two-dimensional configuration space, i.e.~for which the state of the system is determined by two integers $(x,y)$. These two numbers could represent, for example, the state of a biopolymer assembled from two types of monomers X and Y, or from monomers of a single type X but which can be modified (e.g.~via phosphorylation). They could also represent two types of modifications applied to the monomers that make up a fixed-size structure such as a protein complex. Implementing transitions between contiguous $(x,y)$ states results in a lattice-like description of the system. Such a lattice will have boundaries or ``edges'' representing the physical constraints in the system, for example $0 \leq x \leq N_x$ and $0 \leq y \leq N_y$ where $N_x$ and $N_y$ represent e.g.~the number of X and Y monomers available for binding, or the number of binding sites for X and Y in a protein complex. More elaborate constraints can also arise, such as $0 \leq y \leq x$ if $y$ describes the number of monomers in a biopolymer that have undergone some modification out of a total of $x$.

A simple implementation of microscopic out-of-equilibrium ``futile'' cycles, reminiscent of cyclotron orbits in the quantum Hall effect [Fig.~\ref{fig1}(a)], can be achieved in a system with four internal states (A,B,C,D) and four \emph{external} transitions
\begin{eqnarray}
(x,y)_\mathrm{A} \xrightarrow{\gex}  (x,y+1)_\mathrm{B} \nonumber \\
(x,y)_\mathrm{B} \xrightarrow{\gex}  (x+1,y)_\mathrm{C} \nonumber \\
(x,y)_\mathrm{C} \xrightarrow{\gex}  (x,y-1)_\mathrm{D} \nonumber \\
(x,y)_\mathrm{D} \xrightarrow{\gex}  (x-1,y)_\mathrm{A} \nonumber
\end{eqnarray}
which naturally lead to closed cycles $(x,y)_\mathrm{A} \xrightarrow{\gex} (x,y+1)_\mathrm{B} \xrightarrow{\gex} (x+1,y+1)_\mathrm{C} \xrightarrow{\gex} (x+1,y)_\mathrm{D} \xrightarrow{\gex} (x,y)_\mathrm{A}$. We have defined these transitions such that cycles are clockwise in $(x,y)$ space, without loss of generality (reversing all arrows would give counter-clockwise cycles). In the absence of any other transitions, the system will be trapped in such cycles and will not explore the available configuration space. However, the system can break out of a cycle if internal states undergo decay, with four \emph{internal} transitions
\begin{eqnarray}
(x,y)_\mathrm{A} \xrightarrow{\gin}  (x,y)_\mathrm{D} \nonumber\\
(x,y)_\mathrm{D} \xrightarrow{\gin}  (x,y)_\mathrm{C} \nonumber\\
(x,y)_\mathrm{C} \xrightarrow{\gin}  (x,y)_\mathrm{B} \nonumber\\
(x,y)_\mathrm{B} \xrightarrow{\gin}  (x,y)_\mathrm{A} \nonumber
\end{eqnarray}
which enable effective diffusion over the bulk of $(x,y)$ space over time. A possible implementation of these 8 transitions for a biopolymer is shown in Fig.~\ref{fig1}(b). The resulting lattice can be embedded in the plane, as shown in Fig.~\ref{fig1}(c).

 \begin{figure}
 \includegraphics[width=1\linewidth]{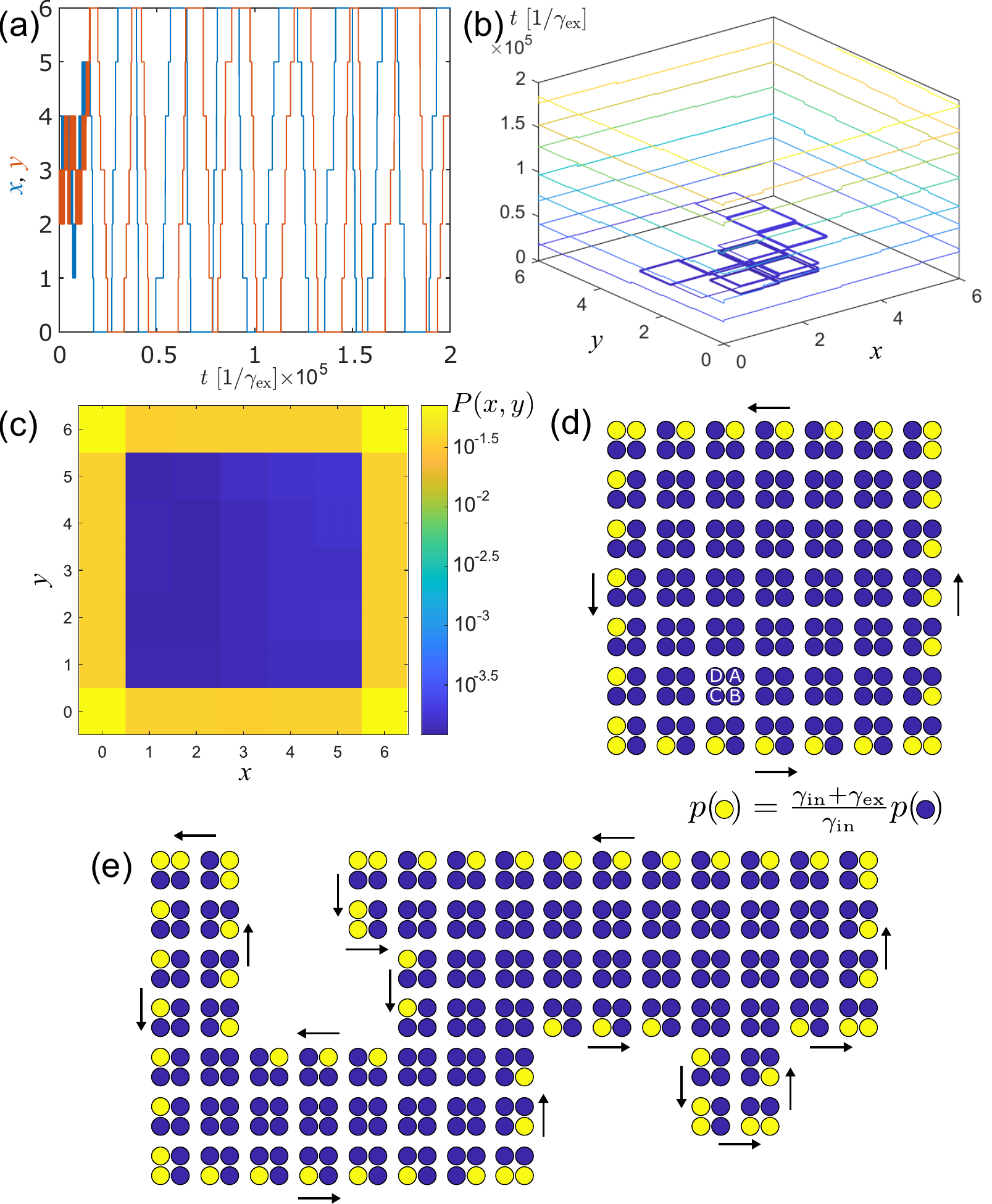}
 \caption{Emergence of global cycles. (a) Simulated stochastic trajectory for the 4-state model [Fig.~\ref{fig1}(c)]. Initially, motion is diffusive, until the system encounters the edge at $y=6$, after which it shows persistent oscillations in both $x$ and $y$. That the oscillations correspond to counter-clockwise edge currents in $(x,y)$ space is clearly seen in (b) which depicts the same trajectory, but in two dimensions (color represents time). (c) The probability distribution in $(x,y)$ space, obtained from simulations, and (d) the steady-state probability distribution in full configuration space, obtained from direct solution of the master equation, both show strong accumulation of probability at the edges. The four arrows in (d) indicate the direction of the edge currents. (e) Edge currents and the resulting cycles are robust with respect to the shape of the boundaries. Parameters used: in (a--c) $\gin=10^{-3} \gex$, in (a--d) system size $N_x=N_y=6$, in (c) $10^7$ steps starting from a random initial state were used. See also Movie 1  in the Supplemental Material \cite{suppmat}. \label{fig2}}
 \end{figure}
Cycles with only three internal states are possible if we allow for diagonal transitions in $(x,y)$, such as $(x,y)_\mathrm{C} \xrightarrow{\gex} (x-1,y-1)_\mathrm{A}$. This gives cycles with only three external transitions, such as $(x,y)_\mathrm{A} \xrightarrow{\gex} (x,y+1)_\mathrm{B} \xrightarrow{\gex} (x+1,y+1)_\mathrm{C} \xrightarrow{\gex} (x,y)_\mathrm{A}$. Including three internal decay transitions for the same reason as above (see Fig.~\ref{fig1}(d) for a possible implementation of the 6 resulting transitions for a biopolymer), we again build a lattice that can be embedded in the plane, e.g.~as a Kagome lattice [Fig.~\ref{fig1}(e)].

Note that, except where explicitly stated, we will consider rectangular geometries in the 4-state model, corresponding to the constraints $0 \leq x \leq N_x$ and $0 \leq y \leq N_y$ and thus a lattice with a total number of $4(N_x+1)(N_y+1)$ states; and triangular geometries in the 3-state model, corresponding to the constraints $0\leq x \leq N_x$ and $0 \leq y \leq x$, which gives a lattice with a total number of $3 (N_x+1)  (N_x+2) / 2$ states. In both cases, the indexing of states starts from $(x,y)=(0,0)$.

\subsection{Chiral edge currents and global cycles \label{sec:edgecurrents}}
 
\begin{figure}[h]
 \includegraphics[width=\linewidth]{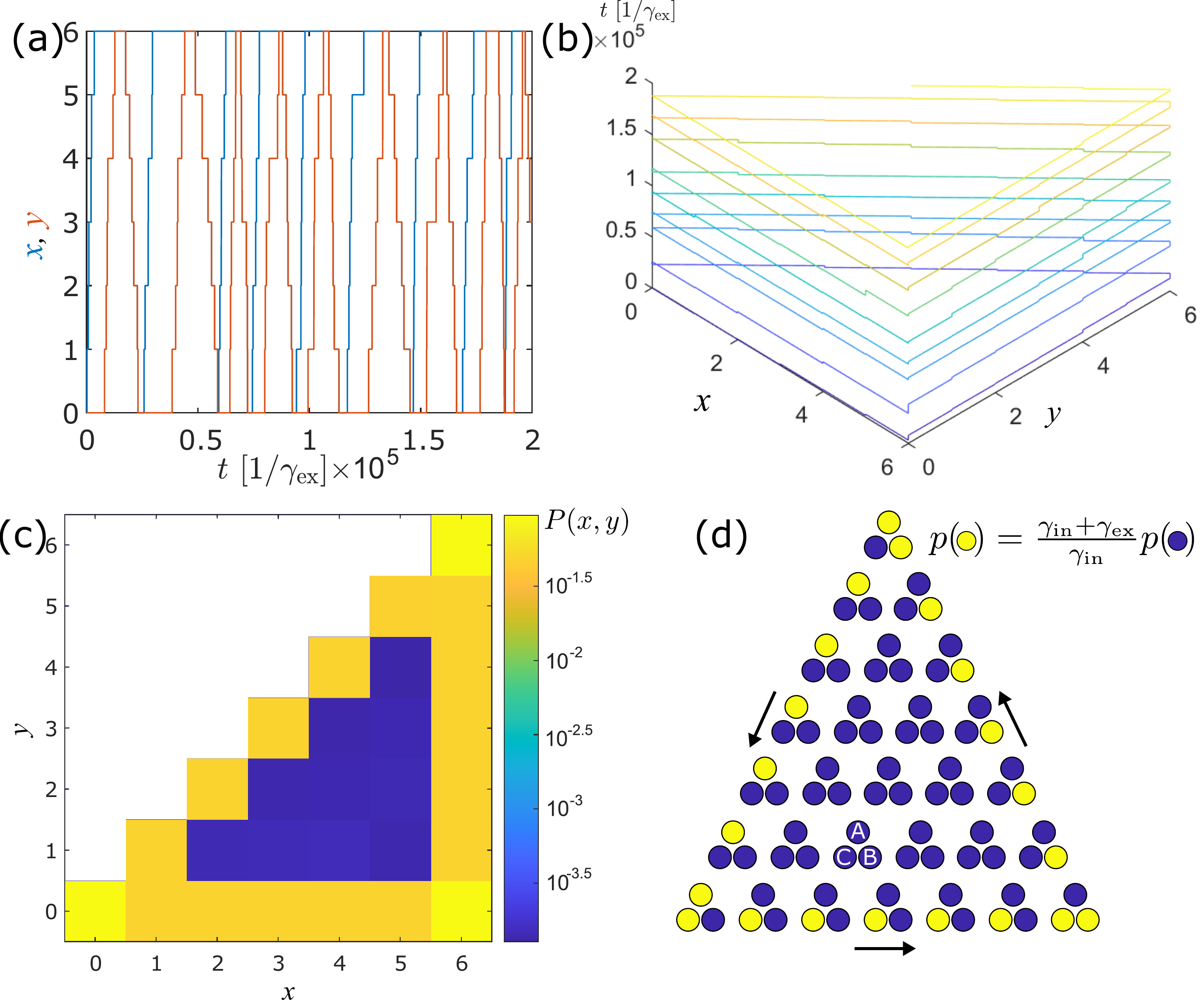}
 \caption{Emergence of global cycles in the 3-state model. (a) Simulated stochastic trajectory for the 3-state model [Fig.~\ref{fig1}(e)] shows persistent oscillations in both $x$ and $y$. (b) Same trajectory, but in two dimensions, clearly showing counter-clockwise cycles (color represents time). (c) The probability distribution in $(x,y)$ space, obtained from simulations, and (d) the steady-state probability distribution in full configuration space, obtained from direct solution of the master equation, both show strong accumulation of probability at the edges. The three arrows in (d) indicate the direction of the edge currents. Parameters used: in (a--c) $\gin=10^{-3} \gex$, in (a--d) system size $N_x=6$ and $y \leq x$, in (c) $10^7$ steps starting from a random initial state were used. See also Movie 2  in the Supplemental Material \cite{suppmat}.  \label{extfig:kagomesim}}
 \end{figure}

Inspection of the lattices in Fig.~\ref{fig1}(c,e) suggests that persistent counter-clockwise trajectories of the system along the edges are possible if $\gex \gg \gin$, i.e.~if the external transition is more likely than the internal one when both are possible (e.g.~at a B state in the bottom edge) so that the system remains on the edge. These trajectories are then analogous to the ``skipping orbits'' in the quantum Hall effect, see Fig.~\ref{fig1}(a). Stochastic simulations of both the 4-state (Fig.~\ref{fig2}) and the 3-state (Fig.~\ref{extfig:kagomesim}) models confirm this expectation. Starting from a state within the bulk of the lattice, the system initially displays local clockwise cycles (driven by $\gex$) interspersed with occasional sideways steps (driven by $\gin$), leading to diffusive motion in the bulk. Once the system reaches any state on the edge, however, persistent motion on the edge leading to counter-clockwise cycles along the boundaries of the system is observed, see Fig.~\ref{fig2}(a,b) and Fig.~\ref{extfig:kagomesim}(a,b). Over time, the probability of finding the system at the edge is significantly larger than in the bulk, see Fig.~\ref{fig2}(c) and Fig.~\ref{extfig:kagomesim}(c), which correspond to histograms obtained directly from the combined $x(t)$ and $y(t)$ trajectories. Direct solution of the steady state probability of the full master equation of the system confirms this result, see Fig.~\ref{fig2}(d) and Fig.~\ref{extfig:kagomesim}(d) and further shows a more detailed structure for the probability of different internal states (or sites) on any given edge cell. This more detailed structured is also recovered from the stochastic simulations (data not shown). As long as $\gex \gg \gin$, the edge cycles are robust to variations in the system size or shape, provided that the directionality of the lattice edges is preserved, see Fig.~\ref{fig2}(e) and Fig.~\ref{extfig:kagome_problem}. We note however that, in the case of Kagome lattices with non-triangular geometry, transitions cutting across concave corners must be absent for edge states to be fully protected, see Appendix \ref{app:stationary1} and Fig.~\ref{extfig:kagome_problem} for more details.

Because the global cycles occur in configuration space, they can describe cyclic changes of a molecular system.  In a system of variable size, such as a biopolymer composed of two types of monomers X and Y and governed by the 4-state model, a global cycle around the system boundary would imply sequential assembly of X monomers, followed by assembly of Y monomers, followed by disassembly of the X monomers, and finally disassembly of the Y monomers, leading back to the initial state. The maximum length in X or Y obtained would only be limited by the availability of monomers. A similar growth-shrinkage cycle can be obtained in the 3-state model, where now $y$ could represent e.g.~the number of monomers that have been dephosphorylated out of a total of $x$. A full cycle would then involve assembly of phospohorylated monomers, dephosphorylation of the monomers, and disassembly of the dephosphorylated monomers. Alternatively, the configuration space could correspond to a system with a fixed number of components, such as a protein complex, but whose components can undergo two types of conformational changes. Such a model, which can describe a biomolecular clock, will be described in Section \ref{sec:globalclock}, where we also explore the period and coherence of the emergent oscillations in more detail.

   \begin{figure*}
 \includegraphics[width=1\linewidth]{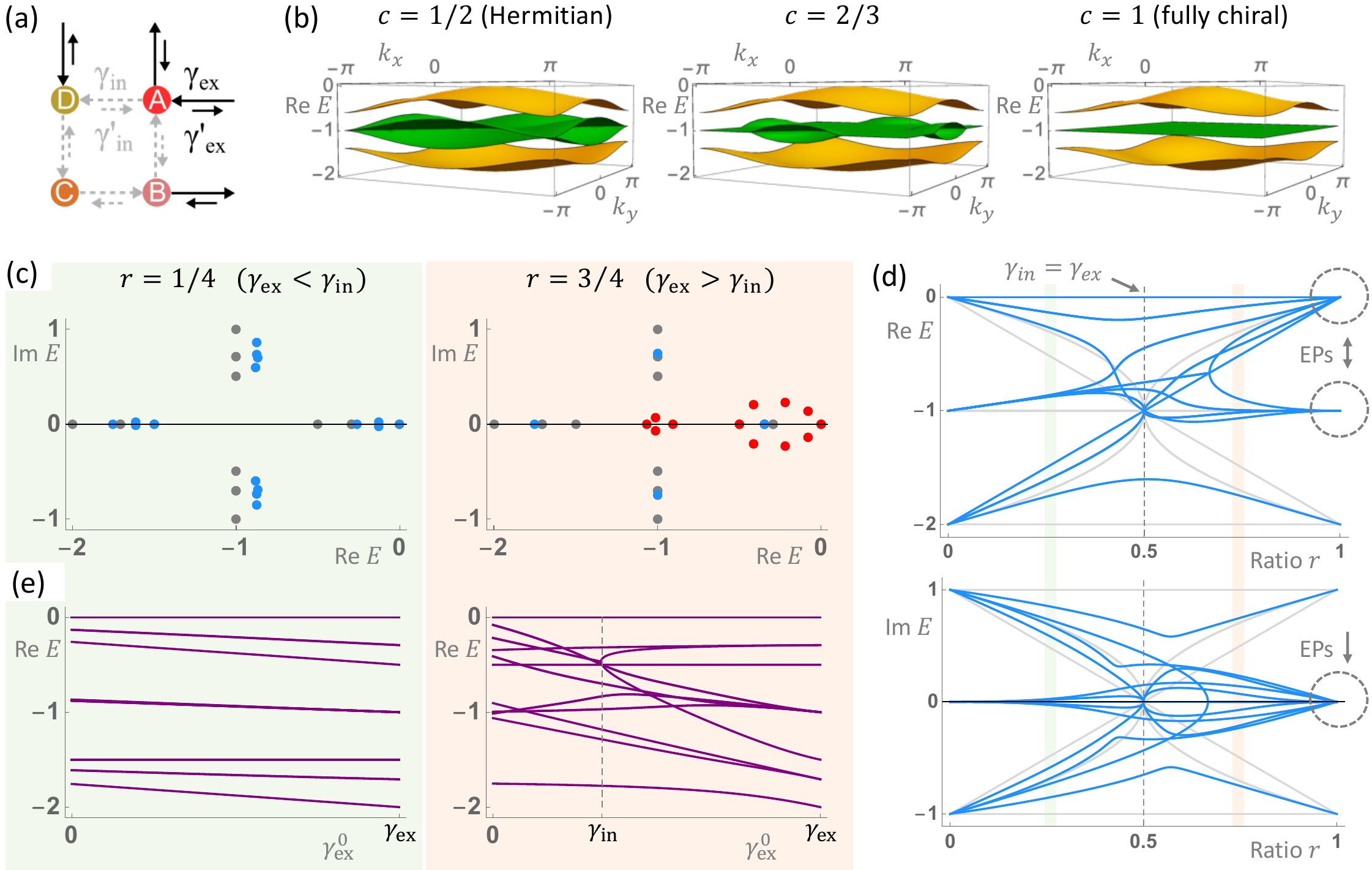}
 \caption{Berry connection and exceptional points. (a) Allowing for the reverse transitions $\gamma_{\textrm{ex}}'$ and $\gamma_{\textrm{in}}'$, we generalize the phase space of our system to be determined by the ratio $r$ weighting external to internal transitions and the chirality $c$, which tunes the system from Hermitian to non-Hermitian. (b) The real spectra of $\mc{W}$ for the square lattice with periodic boundary conditions (PBC) are plotted in reciprocal space $(k_x,k_y)$ for increasing values of  $c$, at $r=0.7$. The top $E_{+,+}$ and bottom $E_{-,+}$ bands are in yellow and the middle bands $E_{+,-}$ and $E_{-,-}$ are in green. The system remains in the topological phase as $c$ increases from 1/2 in the Hermitian case to 1 in the fully chiral limit, since the bandgap remains open. (c) The complex spectra are plotted in the trivial phase ($r=1/4$, green background) and in the Zak phase ($r=3/4$, cream background), where grey denotes periodic boundary conditions (PBC) and blue denotes open boundary conditions (OBC). Edge states (red points) are seen under OBC in the Zak phase, as expected from the bulk-boundary correspondence. (d) As $r\to 1$ in the topological phase, the edge states coalesce towards $E=0$ and $E=-\gamma_\mathrm{tot}$, such that many edge states become close to the ground state in the former. These points of  coalescence are exceptional points (EPs, indicated by dashed circles), unique non-Hermitian features at which the system becomes singular, and are present in OBC (blue) but absent in PBC (grey). (e) Real spectra as a function of the edge link $\gamma_{\textrm{ex}}^0$ that interpolates between PBC ($\gamma_{\textrm{ex}}^0=\gamma_{\textrm{ex}}$) and OBC  ($\gamma_{\textrm{ex}}^0=0$), where EPs emerge in the Zak phase ($r=3/4$, cream background) at $\gamma_{\textrm{ex}}^0=\gamma_{\textrm{in}}$. In contrast, EPs are absent in the trivial phase ($r=3/4$, green background). Parameters used in (c--e): $c=1$ and system size $N_x=N_y=1$. In all panels, values of $E$ are given in units of $\gamma_\mathrm{tot}$.
\label{fig:topo}}
 \end{figure*}

The persistence of edge trajectories can be understood quantitatively. The probability of remaining $L$ steps along the edge and then ``unbinding'' from it is given by $P(L)=\left( 1 - \frac{\gin}{\gin+\gex} \right)^L \frac{\gin}{\gin+\gex}$, which results in an average run length $\langle L \rangle = \sum_{L=0}^\infty L P(L) = \gex/\gin$. Thus, for $\gex = 10^3 \gin$ and $N_x=N_y=6$ as in Fig.~\ref{fig2}, we expect the system to perform $10^3/(6 \cdot 4) \approx 42$ full cycles on average before unbinding. Even then, the system is likely to encounter the edge again soon after and thus undergo a new run along the edge. Moreover, we can analytically obtain (see Appendix~\ref{app:stationary1}) the stationary probability distribution of the system, both in the 4-state and the 3-state models, and find that probability accumulates in the edge sites that precede an internal transition [e.g.~C sites at the bottom edge; see Fig.~\ref{fig1}(c,e)], which have stationary probability $p_C =  \frac{\gin + \gex}{\gin} \,p_b$, where $p_b$ is the probability corresponding to all other sites, including bulk sites as well as edge sites that precede an external transition (e.g.~B sites at the bottom edge). This corresponds to the results in Fig.~\ref{fig2}(d) and Fig.~\ref{extfig:kagomesim}(d), as well as Fig.~\ref{fig2}(e) and Fig.~\ref{extfig:kagome_problem}(b). Importantly, these results imply that probability is \emph{infinitely} localized at the edge sites in the steady state, in the sense that there is not a gradual decay to the bulk probability as we move away from the edge sites. Summing up the probability of all edge sites, we can obtain the overall probability $P_\mathrm{edge}$ of finding the system at the edge at any time, or equivalently, the fraction of time that the system spends at the edge. For a square 4-state system of size $N_x=N_y=N$, in the limit $\gex \gg \gin$, we find $P_\mathrm{edge} \simeq \frac{\gex/\gin}{N+\gex/\gin}$ (see Appendix~\ref{app:stationary1}). In the example of Fig.~\ref{fig2}, this implies that the system spends $\approx 99.4 \%$ of the time at the edge.

 \section{Topological protection of the edge states \label{sec:topology}}
 
 \subsection{Berry connection and transition to the Zak phase \label{sec:berry}}
 The emergence of edge states that dominate the system dynamics is a hallmark of a topological phase, which we can see upon analyzing the Master equation that describes stochastic systems, $\frac{d}{dt}\vect{p}=\mc{W}\vect{p}$. Here $\vect{p}$ is a vector of the probabilities of being in each state, and $\mc{W}$ is a real matrix specifying the transition rates \cite{RevModPhys.48.571}. The lattice structure of the bulk transitions in $\mc{W}$ allows for the calculation of the Berry connection of $\mc{W}$, which determines the topological index of the system, i.e.~whether the system is in the Zak phase. When this index is 0, the system is in the trivial phase, in which case the system will mostly remain in the bulk. In contrast, when the index is $\pi$, long-lived edge states will appear. This relation between the bulk topological phase and its edge properties constitutes the celebrated bulk-boundary correspondence \cite{Cohen2019}.  
 
The link between a particle moving in a lattice and its Berry connection was first formulated by Zak for electrons in a crystal lattice \cite{PhysRevLett.62.2747}. Later, this link was extended to many other systems including nonelectronic ones such as photonics \cite{doi:10.1021/acsphotonics.8b00117,Benalcazar61,PhysRevX.4.021017}, mechanics \cite{Serra-Garcia2018}, acoustics \cite{PhysRevLett.119.255901}, and electrical circuits \cite{Imhof2018}. While the lattice coordinates represent real space in these previous works, in ours they represent configuration space. Remarkably, $\mc{W}$ in our system is a 2d non-Hermitian generalization of the Su-Schrieffer-Heeger (SSH) model for polyacetylene \cite{PhysRevLett.42.1698}, up to a diagonal matrix (see Appendix~\ref{app:symmetries}). When the lattice has periodic boundary conditions (PBC) this diagonal matrix is proportional to the identity matrix, and the eigenvectors of $\mc{W}$ are exactly those of a 2d SSH model, hence they  have the same Berry connection (see Appendix~\ref{app:berry}).  Integration of the Berry connection over reciprocal space is quantized in our system due to the presence of inversion and sublattice (or chiral) symmetries \cite{doi:10.1021/acsphotonics.8b00117,PhysRevB.100.075437,PhysRevLett.62.2747,PhysRevB.97.045106}, and can support the Zak topological phase.

The 2d SSH model on the square \cite{PhysRevB.100.075437} and kagome \cite{Ni_2017} lattices have been previously studied in the Hermitian limit, where the system is in the Zak phase when $\gex>\gin$ and exhibits edge states.
To understand how these properties extend into the non-Hermitian case which characterizes our out-of-equilibrium system, we generalize the phase space. We introduce transitions in the reverse direction from $\gamma_{\textrm{ex}}$ and $\gamma_{\textrm{in}}$, which we call $\gamma_{\textrm{ex}}'$ and $\gamma_{\textrm{in}}'$ respectively [Fig.~\ref{fig:topo}(a)]. In Fourier space, the spectrum of $\mc{W}$ for the 4-state model takes the form (see Appendix~\ref{app:symmetries})
\begin{eqnarray}
E(\vect{k})_{\pm,\pm}= -\gamma_{\textrm{tot}}\pm\sqrt{a(\vect{k})\pm\sqrt{a(\vect{k})^2-b(\vect{k})}}\label{eq:periodicspect}
\end{eqnarray}
where $\gamma_\mathrm{tot} \equiv \gin+\gpin+\gex+\gpex$ and  
\begin{eqnarray}
a(\vect{k})&=&2 (\gin\gpin + \gex\gpex) + (\gin\gex + \gpin\gpex)(\cos k_x + \cos k_y)\nonumber \\
b(\vect{k})&=&\left( \beta-\beta' + 2\gpex\gin\cos k_x - 2\gex\gpin\cos k_y\right)\nonumber \\&&(\beta'-\beta+ 2 \gex\gpin\cos k_x - 2\gpex\gin \cos k_y)\nonumber \\
\beta&=&\gin^2-\gex^2,\qquad\beta'=\gpin^2-\gpex^2.
\end{eqnarray}

To simplify notation, we introduce two parameters, the ratio $r$ and the chirality $c$, where $\gin = c (1-r) \gamma_\mathrm{tot}$, $\gpin = (1-c)(1-r)\gamma_\mathrm{tot}$, $\gex = c r \gamma_\mathrm{tot}$ and $\gpex = (1-c) r \gamma_\mathrm{tot}$. The ratio $r$ weights the relative strength of internal and external transitions, with $r>1/2$ ($r<1/2$) when external (internal) transitions are stronger. The chirality parameter $c$ weights the relative strength of forward and reverse transitions, where we note that $\gin + \gex = c
\gamma_\mathrm{tot}$ and $\gpin + \gpex = (1-c)\gamma_\mathrm{tot}$. For $c>1/2$, external and internal cycles are biased in the clockwise and counter-clockwise direction, respectively, and edge currents are counter-clockwise. For $c<1/2$, the chiralities are opposite, with external and internal cycles biased in the counter-clockwise and clockwise direction, respectively, and edge currents being clockwise. The value $c=1/2$ corresponds to the Hermitian, achiral case with equal forward and reverse rates for all transitions. Finally, $c=1$ describes the fully chiral case with only forward rates studied above (see Fig.~\ref{fig1}), and $c=0$ is also fully chiral but with opposite chirality.

  \begin{figure*}
 \includegraphics[width=0.85\linewidth]{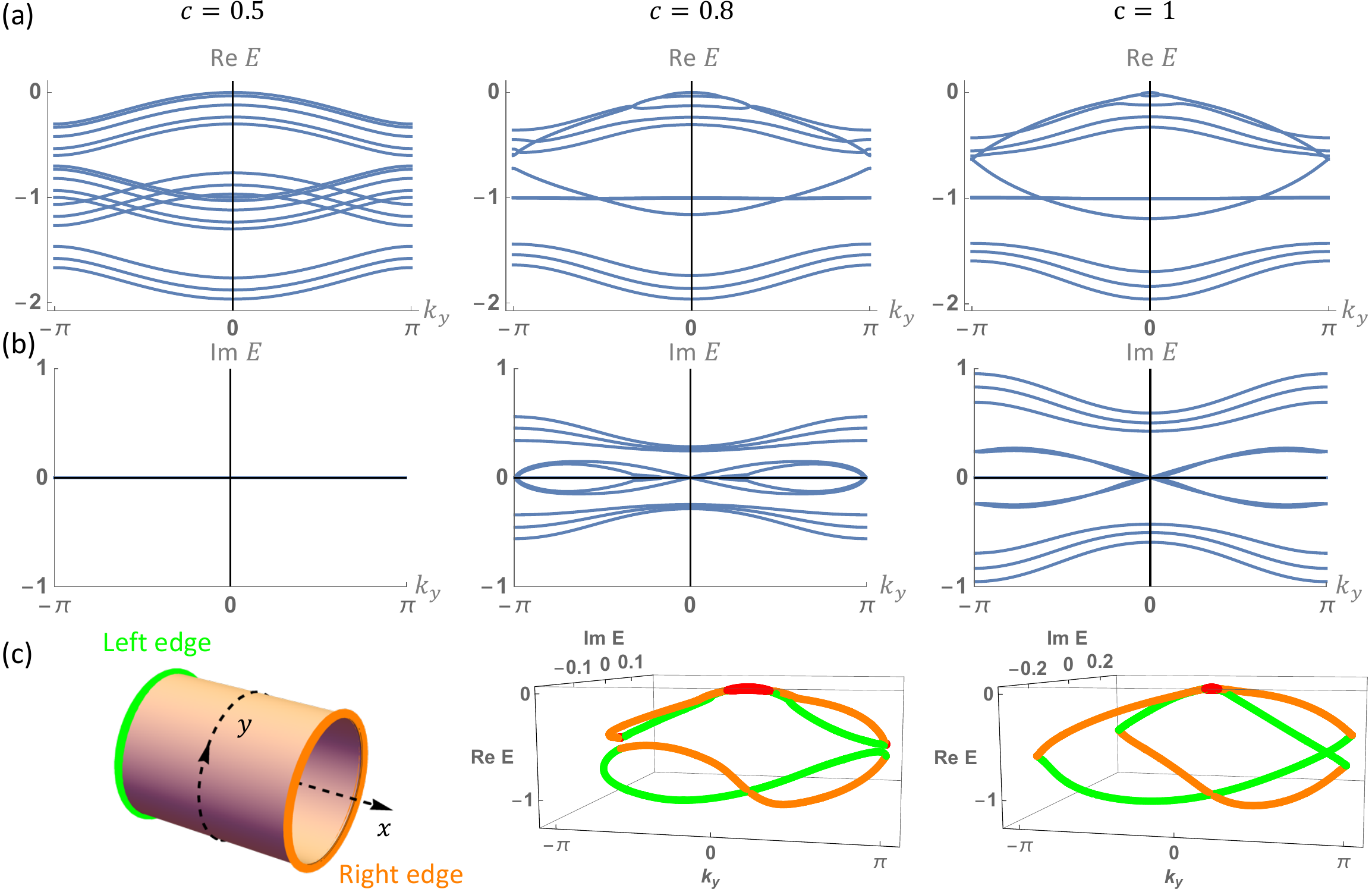}
 \caption{Emergence of edge states and the non-Hermitian topological invariant, vorticity. (a) Using a half-periodic geometry with  OBC in $x$ and PBC in $y$, we calculate the band structure of the system along the reciprocal lattice index $k_y$. In the Zak phase at $r=0.7$, we show three examples as the system is varied from the Hermitian case with equal forward and reverse transitions ($c=0.5$) to the full chiral case with only forward transitions at $c=1$. \textit{Left}: In the Hermitian limit, all the bands have similar amount of group velocity $d (\textrm{Re}E)/d k_y$ or bandwidth. \textit{Center and right}: As $c$ increases, bands localized on the edge emerge, which are the bands with the largest group velocity.  (b) \textit{Left}: When $\mc{W}$ is Hermitian, its bands are completely real. \textit{Center and right}: $\mc{W}$ is non-Hermitian ($c\neq0.5$), its bands have an imaginary component which increases in magnitude with $c$.  (c) \textit{Left}: A schematic of the semi-periodic geometry with different bands localized on the system edge.  \textit{Center and right}: Edge states, i.e.~the two pairs of bands with largest group velocity, in complex space. Part of the bands are completely real (red) and distributed in equal amounts on the left and right edges. As chirality increases (\textit{Center}, $c=0.8$), they increasingly split in complex space and become localized on the left (green) and right (orange) edges respectively, but still exhibit a vorticity of $\nu=0$. For sufficiently strong chirality (\textit{Right}, $c=1$), a transition occurs when the two pairs of edge bands touch in real space, exhibiting a doubled periodicity of $4\pi$ and vorticity of $\nu=1/2$. All panels use $N_x=3$ and give values of $E$ in units of $\gamma_\mathrm{tot}$. \label{extfig:vorticity}}
 \end{figure*}
 
We can now examine if the system remains topological as chirality is tuned from the Hermitian case to the fully chiral case, by verifying if the system bandgap remains open throughout. Using Equation~\ref{eq:periodicspect}, we indeed note that the bandgap remains open as we vary $c$ from 1/2 to 1 above a critical value $r^*$ [Fig.~\ref{fig:topo}(b)]. We find that $r^*=1/2$ for both the purely Hermitian case \cite{PhysRevB.100.075437} ($c=1/2$) and the fully chiral case ($c=1$ and $c=0$), see Appendix~\ref{app:berry} and Fig.~\ref{extfig:periodicbands} there. However, stronger external transitions are needed for the transition in between these values, similar to when finite-size effects come into play \cite{PhysRevB.100.075437}. The value $r^*(c)$ can be obtained from the maximum value of $r$ for which $a(\vect{k})^2=b(\vect{k})$ has a solution. The global maximum across all values of $c$ is $r^*_{\textrm{max}}=0.59$, which occurs at $c=0.7$ (and $c=0.3$). 

In the topological phase, i.e.~when $r>r^*$, the system exhibits edge states in open boundary conditions (OBC) as expected from the bulk-boundary correspondence. These states are colored in red in Fig.~\ref{fig:topo}(c), forming two rings that shrink into points as $r\to1$ . The larger of these rings shrinks towards the ground state at $E=0$, creating a large number of edge states that are increasingly close to the ground state as $r\to1$. Note that the total number of edge states scales with the number of sites on the edge, e.g.~$4(N_x+N_y-1)$ in a rectangular geometry.

 \subsection{Uniquely non-Hermitian features: propagating edge states, exceptional points and vorticity \label{sec:nonhermitian}}
 We find that our system exhibits several uniquely non-Hermitian properties in the chiral case, which is also when the edge  states change qualitatively from being stationary polarization to chiral currents. Such non-Hermitian features include exceptional points (EPs) \cite{Heiss:2012dx,PhysRevLett.116.133903,Xiong_2018,PhysRevLett.123.205502,You19767,PhysRevX.10.041009} or topological invariants without a Hermitian counterpart \cite{PhysRevLett.120.146402,2020arXiv200601837A}. These unique non-Hermitian properties emerge only in the finite Zak phase and most notably in the case of OBC. In the fully chiral case, the spectrum of the system with PBC (grey lines) is symmetric with respect to $r^*=1/2$ (where $\gamma_{\textrm{ex}}=\gamma_{\textrm{in}}$) [Fig.~\ref{fig:topo}(d)]. In OBC (blue lines), the spectrum changes radically past $r^*$, with many states coalescing towards $E=0$ and some towards  $E=-\gamma_\mathrm{tot}$ as $r\to1$, indicating the existence of EPs\cite{Heiss:2012dx,Xiong_2018} at $r=1$ (see dashed circles in Fig.~\ref{fig:topo}(d)). The transition at $r=r^*$ coincides with when the system is just as likely to unbind as to remain on the edge at every step, with the average run length of $\<L\>=1$, whereas the limit of $r\to1$ corresponds to when the system spends all of its time on the edge. Lastly, varying the edge transition rate $\gamma_{\textrm{ex}}^0$ (see Appendix~\ref{app:symmetries}) interpolates between PBC ($\gamma_{\textrm{ex}}^0=\gamma_{\textrm{ex}}$) and OBC ($\gamma_{\textrm{ex}}^0=0$). In the Zak phase, exceptional points emerge in the spectrum at $\gamma_{\textrm{ex}}^0=\gamma_{\textrm{in}}$ [Fig.~\ref{fig:topo}(e)].  
 



Another uniquely non-Hermitian feature emerges in the edge states in the strongly chiral limit, that of vorticity \cite{PhysRevLett.120.146402}. Using a half-periodic geometry with OBC in $x$ and PBC in $y$ (Fig.~\ref{extfig:vorticity}), we can  calculate the band structure (eigenvalues) of $\mc{W}$ along the reciprocal lattice index $k_y$. In the Hermitian limit of $c=1/2$, all the bands have similar amount of group velocity $d (\textrm{Re}E)/d k_y$ and are completely real, see Fig.~\ref{extfig:vorticity}(a,b) (left). When Hermiticity is broken and the system becomes increasingly chiral, bands localized on the edge emerge, which also exhibit the largest group velocity, see Fig.~\ref{extfig:vorticity}(a) (center and right). Many bands also have an imaginary component which increases in magnitude with $c$, see Fig.~\ref{extfig:vorticity}(b) (center and right). As the group velocity describes the propagation speed of probability disturbances, the increase of group velocity with chirality implies a decrease in the period of oscillations around the edge of the system with increasing chirality, which is supported by the explicit calculations in Section \ref{sec:globalclock}. 

These edge states demonstrate a topological transition with increasing chirality. Part of these bands are completely real (red) and distributed in equal amounts on the left and right edges [red in Fig.~\ref{extfig:vorticity}(c)]. As chirality increases, they develop growing imaginary components to become localized on the left (green) and right (orange) edges respectively. In the strongly non-Hermitian limit, a transition occurs when the two pairs of edge bands touch in real space, exhibiting a doubled periodicity of $4\pi$ and vorticity of $\nu=1/2$.  Vorticity is a uniquely non-Hermitian topological invariant describing the winding number of a pair of bands in the complex plane \cite{PhysRevLett.120.146402}: 
\begin{eqnarray}
\nu_{mn}(\Gamma)=-\frac{1}{2\pi}\oint_{\Gamma}\vect{\nabla}_{\vect{k}} \textrm{arg}[E_m(\vect{k})-E_n(\vect{k})]\cdot d\vect{k},\label{eq:topoinv}
\end{eqnarray}
where $\Gamma$ is a closed loop in reciprocal space, and $m,n$ are band indices. In our system, $\nu$ indicates the strongly propagating nature of the edge states, taking the value of 0 at $c=0.8$  and 1/2 at $c=1$. 

\begin{figure*}
 \includegraphics[width=1\linewidth]{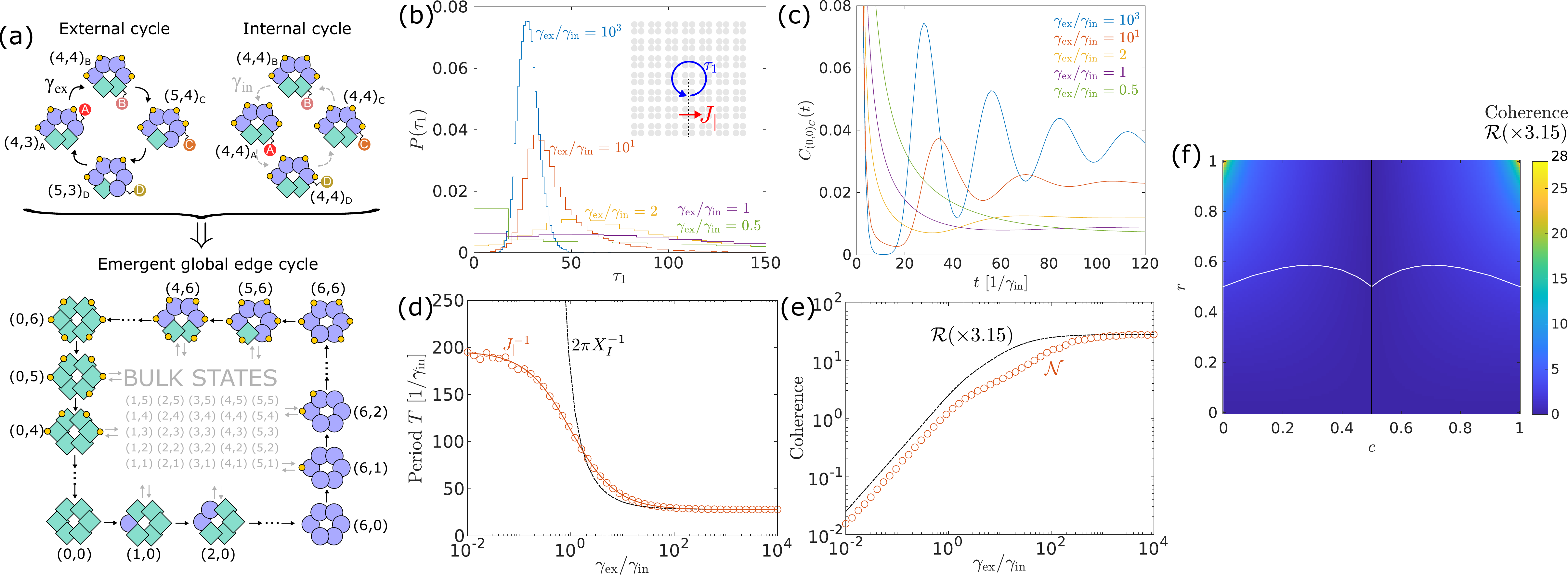}
 \caption{ (a) 4-state model (with $N_x=N_y=6$) for a hexameric biomolecular oscillator such as the KaiABC system, involving allosteric conformational changes of the monomers (circles to squares) and phosphorylation (yellow added circles). The emerging topologically-protected edge state of this model corresponds to cycles of conformational change, phosphorylation, conformational change, dephosphorylation. (b) Probability distribution of times $\tau_1$ to complete one cycle obtained from stochastic simulations, defined as two subsequent crossings of the dashed line shown in the inset (the flux through which is denoted as $J_|$), for various values of $\gex/\gin$. (c) Correlation function representing the probability of finding the system in state $(0,0)_C$ at time $t$ given that it was in said state at time zero, for the same values of $\gex/\gin$, showing damped oscillations. (d) Period $T$ of the oscillations as a function of $\gex/\gin$, obtained from (circles) the mean cycle time in stochastic simulations, (red solid line) the inverse of the analytically-calculated steady-state flux $J_|$, and (dashed black line) the inverse of the imaginary part $X_I$ of the slowest decaying eigenmode of the transition matrix $\mathcal{W}$. (e) Two measures, $\mathcal{N}$ (obtained from stochastic simulations) and $\mathcal{R}$ (obtained from the slowest decaying eigenmode of $\mathcal{W}$), for the coherence of the oscillations as a function of $\gex/\gin$, see the main text for their definition. (f) Coherence $\mathcal{R}$ in the full parameter space spanned by $r$ and $c$, showing that coherence is maximized in the fully chiral limit ($c \to 0$ or $1$) and deep in the topological phase ($r \to 1$). The white curve represents the boundary $r^*(c)$ separating the topological and trivial phases of the system. For each curve in (b) and each data point in (d,e) we used $10^7$ stochastic simulation steps starting from a random initial state.   \label{fig:oscillations}}
 \end{figure*}

To summarize the topological properties of the system, we find that (i) topological edge states occur when $r > r^*(c)$ (the shape of this boundary can be seen in Figs.~\ref{fig:oscillations}(f) and \ref{extfig:fluxesphasediag}); (ii) when $c=0.5$, the system is Hermitian and no chiral probability currents are observed at the boundary; and (iii) for $c>0.5$ probability currents at the
edge emerge in a counter-clockwise direction; whereas for $c<0.5$ currents emerge in the clockwise direction. Topologically, the emergence of protected chiral edge currents is manifested in the transition from zero vorticity $\nu=0$ to non-zero vorticity $\nu=1/2$ and $\nu=-1/2$ as chirality increases or decreases away from $c=0.5$, respectively. 

 \subsection{Edge localization in steady-state \label{sec:localization}}

Properties of the non-Hermitian system can also be analyzed using a transfer matrix approach, which probes the steady state of the full transition rate matrix $\mc{W}$ (see Appendix~\ref{app:stationary2}), in the general case of arbitrary $c$ and $r$. This yields the probability accumulation and current along the edge in the full phase space of $c$ and $r$, see Fig.~\ref{extfig:fluxesphasediag} in the Appendix. In particular, we find that both the probability accumulation and the current vanish in the achiral, Hermitian case with $c=0.5$, as well as in the limit $r \to 0$. The current also vanishes in the limit $r \to 1$, at which point the sites along the edge become disconnected because $\gin,\gpin \to 0$.

Moreover, the transfer matrix analysis shows that the edge states are exponentially localized, in the sense that the probability accumulation with respect to the bulk probability in a cell situated $n$ cells away from the edge $\delta P_n$ decays as $\delta P_n = \delta P_0 \alpha^n$, where $\alpha$ is a decay constant satisfying $0 \leq \alpha < 1$ that depends on both $c$ and $r$. Importantly, in the limit of a fully chiral system with $c \to 1$ or $c \to 0$ we find $\alpha = 0$, and thus recover the result of infinite localization at the edges described in Sec.~\ref{sec:edgecurrents}.

  \subsection{Identifying topology in realistic systems \label{sec:realistic}}
While some of the results above are obtained in a periodic setting, i.e.~representing an infinite system without disorder, they are directly applicable to realistic systems which are typically finite and may have some amount of disorder in the transition rates. Thanks to the bulk-boundary correspondence \cite{Cohen2019,CHEN2020126168}, our results for the topology of $\mc{W}$ in a periodic setting will manifest as edge state properties when a finite system is considered, or when an edge is introduced in an otherwise open system. Further, since our edge states exponentially decay into the bulk as discussed in the previous section, finite size effects would not alter the bulk properties in systems that are larger than the exponential decay length. In practice, this allows the bulk-boundary correspondence to hold even for small systems with just a few unit cells, for instance in the fully chiral limit where the edge states are infinitely localized. This permits the calculation of the topology of $\mc{W}$ for smaller systems, where we expect an edge state at the boundary of the system in the topological phase, or at the boundary between two subsystems that are in topological and trivial phases respectively.

The bulk-boundary correspondence also remains robust to weak disorder, as has been found from both analytical \cite{CHEN2020126168} and numerical \cite{PhysRevLett.95.136602} calculations. Specifically, this is true when the disorder does not does not cause the bandgap to close, which preserves the topological phase \cite{PhysRevLett.95.136602}, e.g.~if disordered transitions are too slow to cause scattering between the edge and bulk states. This can be checked by direct calculation of the disordered bulk $\mc{W}$. Alternatively, inspection of the eigenstates of $\mc{W}$ in a finite geometry in the presence of disorder can reveal whether the long-lived states remain on the edge or have merged with the bulk states. The presence of strong disorder in such systems merits systematic exploration in future work.

 \section{Applications to soft and living matter \label{sec:complex}}
 
 \subsection{Global clock \label{sec:globalclock}}
 
 As anticipated in Section~\ref{sec:edgecurrents}, the two-dimensional configuration spaces described here may represent a protein complex with a fixed number of components that can undergo two types of conformational changes. As an example, in Fig.~\ref{fig:oscillations}(a) we show how our 4-state model with $N_x=N_y=6$ would support the dynamics of an allosteric model for a hexameric biochemical oscillator such as the KaiABC system. Typical models for such conformational oscillations \cite{vanZon7420,Chang2012,Zhang2020} rely on the concerted or Monod-Wyman-Changeux paradigm \cite{pmid14343300} of allosteric regulation which restricts the configuration space to states in which either all or none of the monomers in a complex have undergone a conformational change. The latter traditionally stands in opposition with the sequential or Koshland-Némethy-Filmer paradigm \cite{koshland1966}, which allows individual monomers to separately undergo conformational changes. Interestingly, our system shares features of both paradigms: while all states in the full two-dimensional configuration space are allowed, as in the sequential model, the system only visits a strongly limited subset of these states, namely those at the edge, as in the concerted model.  Moreover, this property arises from the repetition of simple, local reaction motifs without the need of fine-tuning, e.g.~of some transitions to be markedly different from others at particular points in configuration space.

 In order to obtain a functional clock, it is important that oscillations have a reproducible period. We have characterized the period of oscillations for our 4-state model with $N_x = N_y = 6$ in the fully chiral case ($c=1$, i.e.~$\gpex=\gpin=0$), as a function of the ratio $\gex/\gin$. \rev{Note that $
\gex/\gin = r/(1-r)$, where $r$ is the ratio parameter introduced in the previous Section~\ref{sec:berry}.} In Fig.~\ref{fig:oscillations}(b), we show the distribution of the times $\tau_1$ needed to complete a full cycle, defined as the time between two consecutive crossings of a reference line which is marked as the dashed line in the inset of Fig.~\ref{fig:oscillations}(b), obtained from stochastic simulations, for several values of $\gex/\gin$. In the topological regime with $\gex/\gin \gg 1$, we find a strongly peaked distribution, implying a well-defined oscillation period. The average oscillation period can be taken to be $T \equiv \langle \tau_1 \rangle = J_|^{-1}$, where $J_|$ is the average current through the same line that is used to define the cycle time. As the ratio $\gex/\gin$ is decreased, this distribution broadens until, for $\gex/\gin \lesssim 1$, the distribution is no longer peaked and instead shows a standard deviation comparable to or larger than its mean. From the mean and variance of the cycle time we can define a measure of coherence $\mathcal{N}  \equiv  \langle \tau_1 \rangle^2 / \mathrm{var}(\tau_1)$ (inverse to the Fano factor) \cite{barato2017coherence,marsland2019thermodynamic}, which represents the average number of oscillations that the system will undergo before an error of plus or minus one oscillation has accumulated and coherence is lost. The period and coherence obtained from stochastic simulations in this way are shown as the red circles in Figs.~\ref{fig:oscillations}(d) and (e), respectively. Moreover, we can calculate the average current $J_|$ in steady state explicitly for a square 4-state model with $N_x=N_y=N$, see Appendix \ref{app:stationary1}, and thus the oscillation period as
 \begin{equation}
     T = J_|^{-1} = \frac{4(N+1)}{\gin} \cdot \frac{\gex+(N+1)\gin}{\gex+\gin}
     \label{eq:T}
 \end{equation}
 which is plotted as the red solid line in Fig.~\ref{fig:oscillations}(d) and matches the simulation results. In the limit $\gex/\gin \gg N$, the period tends to $T=4(N+1)/\gin$ (or $T=28/\gin$ for $N=6$), which represents the fact that in this limit the edge states become isolated from the bulk, the waiting times for external transitions become negligible with respect to those for internal transitions, and the system thus becomes equivalent to a unidirectional cycle with transition rate $\gin$ between $4(N+1)=28$ states.
 
 An alternative way to define the period and coherence of oscillations, which bypasses the need for stochastic simulations, is to consider the time correlation function of a given state, which can be obtained directly at the level of the master equation for the probabilities $\frac{d}{dt}\vect{p}=\mc{W}\vect{p}$. \cite{barato2017coherence} In Fig.~\ref{fig:oscillations}(c), we show the correlation function $C_{(0,0)_C}(t)$, i.e.~the probability of finding the system in state $(0,0)_C$ (which corresponds to the bottom left corner of the system, see Fig.~\ref{fig1}(c)) at time $t$ given that the system was in that state at time zero, again for several values of $\gex/\gin$. The correlation function shows a characteristic behavior, with damped oscillations that ultimately asymptote to the steady-state probability of state $(0,0)_C$. At long times, the decay time of the exponential damping as well as the oscillation period are governed by the first non-trivial eigenvalue of the transition matrix $\mc{W}$, that is, the eigenvalue $\lambda = -X_R \pm i X_I $ with smallest non-zero $X_R$, which corresponds to the slowest-decaying mode in the system. The real part $X_R$ corresponds to the inverse of the decay time, whereas the oscillation period $T'$ is related to the imaginary part $X_I$ through $T' = 2\pi/X_I$ \cite{barato2017coherence,qian2000pumped}. The coherence of the oscillations can thus be measured as the number of oscillations that fit within a decay time, and is directly proportional to the ratio $\mathcal{R} \equiv X_I / X_R$. The oscillation period and coherence obtained in this way are plotted as the black dashed lines in Fig.~\ref{fig:oscillations}(d,e).
 
 Interestingly, while in the topological regime $\gex/\gin \gg 1$ we find that both measures for the oscillation period coincide, the two diverge for $\gex/\gin \lesssim 1$, see Fig.~\ref{fig:oscillations}(d), once again pointing to a qualitative change in the nature of the oscillations in the system between the topological and trivial regimes. Indeed, at low $\gex/\gin$, the internal cycle at the center of the bulk of the system starts to contribute significantly to the current $J_|$, thus lowering the average period $T = J_|^{-1}$, which in any case becomes an uninformative quantity given its large standard deviation, see Fig.~\ref{fig:oscillations}(b). On the other hand, the two coherence measures $\mathcal{N}$ and $\mathcal{R}$ show very similar behavior as a function of $\gex/\gin$, see Fig.~\ref{fig:oscillations}(e). Note that the two differ roughly by a proportionality constant: in the limit $\gex/\gin \gg 1$, we can again exploit the fact that our system becomes a unicyclic network with $4(N+1)$ states. For a unicyclic network with $S$ states, it is known \cite{barato2017coherence} that the two coherence measures are related through $\mathcal{N}/\mathcal{R} = S / \cot (\pi / S)$, which for $S=4(N+1)=28$ results in the proportionality constant $\mathcal{N}/\mathcal{R} \approx 3.15$ used in Fig.~\ref{fig:oscillations}(e).
 
 So far, we have focused on the fully chiral case with $c=1$. In Fig.~\ref{fig:oscillations}(f), we show the coherence in the full $(c,r)$ parameter space, again calculated from the first non-trivial eigenvalue of the transition matrix as $\mathcal{R} \equiv X_I / X_R$. As expected, oscillations are most coherent in the limit of full chirality $c \to 1$ (or $c \to 0$) and large $\gex/\gin$, i.e.~$r\to 1$. Conversely, coherence vanishes in the achiral case $c = 0.5$ and in the limit $r \to 0$. In the limit $r\to 1$ ($\gex/\gin \gg 1$), we can once more use the equivalence to a unicyclic network to calculate the oscillation period as $T=4(N+1)/|\gin-\gpin|$ and the coherence as $\mathcal{N} = 4 (N+1) |\gin-\gpin|/(\gin+\gpin)$ \cite{barato2017coherence}. This shows that the maximal coherence obtainable, which occurs in the fully chiral case, is $\mathcal{N} = 4 (N+1)$, and is thus purely limited by the number of states in the system. This results in the maximum value of 28 obtained for the coherence in Figs.~\ref{fig:oscillations}(e,f).
 
 Returning to the comparison between our topological model for the KaiABC clock and the usual concerted model \cite{vanZon7420,Chang2012,Zhang2020}, it is worth noting that, by including the intermediate states that are disallowed in the concerted model, we raise the upper bound that can be achieved for the coherence of oscillations in the clock. Indeed, the concerted model has $7 \times 2 = 14$ states and thus an upper bound of $\mathcal{N} = 14$ for its coherence for the case of a unidirectional cycle \cite{barato2017coherence}. Instead, our topological model has $7 \times 7 \times 4 = 196$ states, but in the optimal regime becomes equivalent to a unidirectional cycle with $4 \times 7 = 28$ states, leading to an upper bound $\mathcal{N} = 28$ for its coherence.

 \begin{figure}
 \includegraphics[width=1\linewidth]{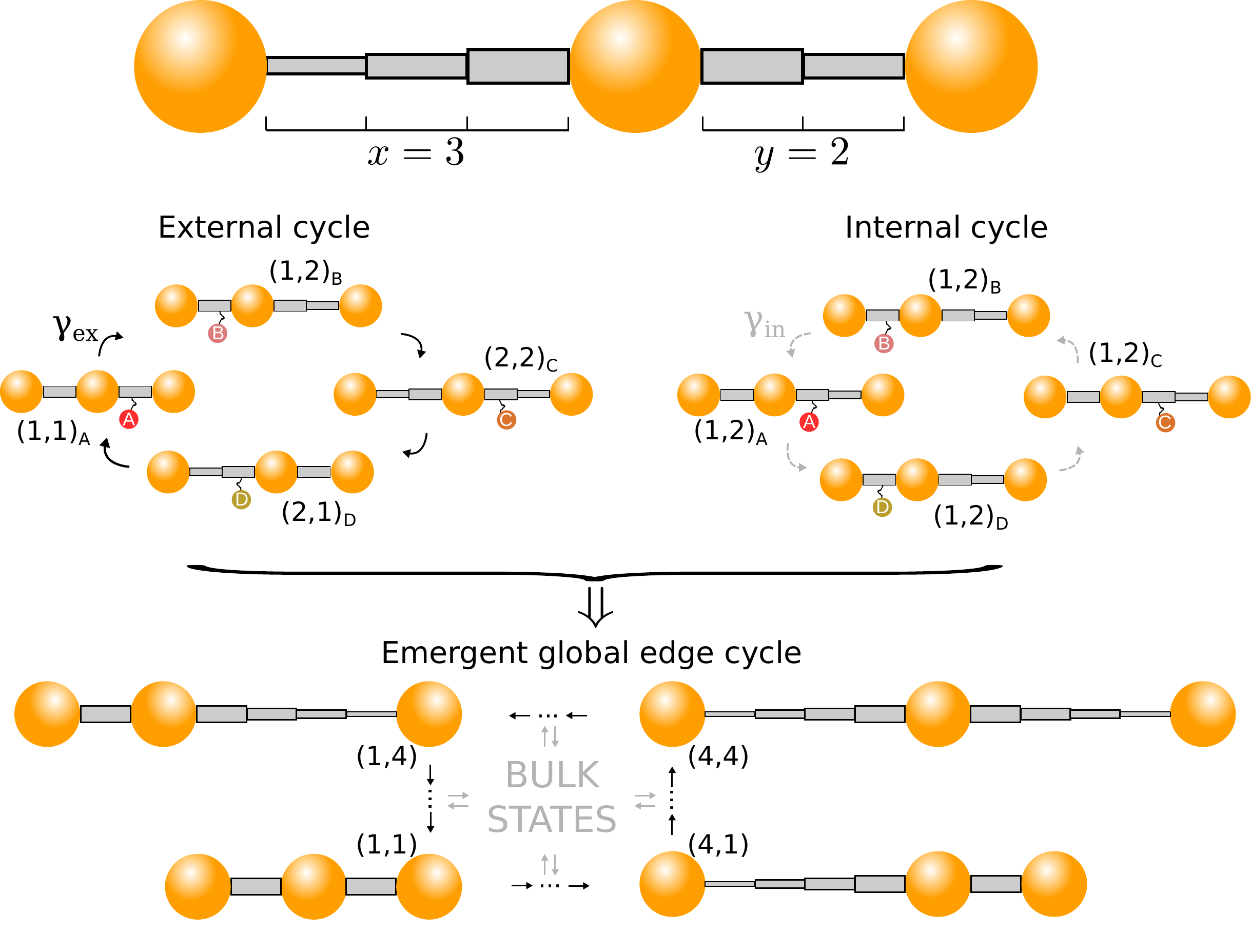}
 \caption{ A topologically-protected three-sphere swimmer, implemented within our 4-state model. In this case, the coordinates $x$ and $y$ describe the extension of the left and right arms, respectively. The internal states A, B, C, D represent whether the left or right arm is primed for increasing or decreasing its length. The emergent topologically-protected edge current results in cyclic shape deformations that enclose the maximally-available area in shape space (in this example, we set the minimal and maximal extension of the arms to be 1 and 4, respectively). \label{fig:swimmer}}
 \end{figure}

 \subsection{Stochastic low Reynolds number swimmers \label{sec:swimmer}}
 
 In another example for a broad class of systems for which the emergence of chiral edge currents in a two-dimensional configuration space is desirable, the  global cycles we predict would constitute an ideal driving mechanism for synthetic nanoscale swimmers \cite{RG2008,RG2010}. At such small scales, hydrodynamic flows occur at low Reynolds number, and the ``scallop theorem'' implies that swimming is only possible if the swimming strokes (shape deformations) exerted by the swimmer are nonreciprocal, thus breaking time-reversal symmetry \cite{purcell1977life}. In particular, the requirement of nonreciprocity implies that the cyclic shape deformations undergone by a low Reynolds number swimmer must be described by at least two parameters, and that these degrees of freedom must describe cycles that enclose a finite area in the (at least two dimensional) shape parameter space. The swimming capabilities of the swimmer directly increase with the area of said cycle.
 
 The topologically-protected edge currents that we predict naturally describe such cycles in a two-dimensional space, and in fact always describe the largest possible cycle within the available parameter space. As a minimal example, in Fig.~\ref{fig:swimmer} we show how to implement a topologically-protected version of the three-sphere swimmer \cite{najafi2004simple} within our 4-state model. The parameters $(x,y)$ correspond to the length of the left and right arms of the swimmer, and we take the minimal and maximal elongation achievable by each arm to be 1 and 4, respectively. Even if all the possible intermediate $(x,y)$ states are accessible, the swimmer spontaneously cycles around the edge states, thus maximizing the area enclosed in parameter space. This cycle emerges without a need for each arm segment to be controlled in the right sequence, reducing the number of external inputs required and/or the internal complexity of the device.
 
The topological protection and emergent nature of the edge states makes the swimmer robust under external perturbations and constraints. For example, even if an internal malfunction, or the presence of an external obstacle, prevented one of the arms from reaching its full length, the system would still spontaneously describe maximally large cycles in the reduced available parameter space. Moreover, even if an external influence or large fluctuation drove the system into a bulk state, which in this context would imply making both arms ``out of step'' with each other, the system will diffuse through the bulk of the state space until it reaches an edge state and resumes its global shape cycles. Such occasional unbinding from the edge states will result in run-and-tumble-like behavior for the swimmer.

 \subsection{Asymmetric rates and dynamic instability \label{sec:asymmetric}}

The inclusion of further features inspired by biology into these models reveals striking observations and directions for future research. For instance, we have until now considered systems characterized by identical rates for the transitions between different states, i.e.~with a single value of $\gex$ and $\gin$ for all the external and internal transitions, respectively. This symmetry need not exist in general, and indeed, in real systems we expect that the transition rates between the different states will be different from each other. We introduce superindices to denote the transition rates between two specific states such that, for example, $\gamma_\mathrm{ex}^{BC}$ is the rate of the external transition from B to C, and $\gamma_\mathrm{in}^{CB}$ of the internal transition from C to B. In general, there are thus 8 transition rates in the fully chiral 4-state model and 6 transition rates in the fully chiral 3-state model. Robust edge currents will survive as long as the external transitions are significantly faster than the internal transitions with which they compete.

\begin{figure}
 \includegraphics[width=1\linewidth]{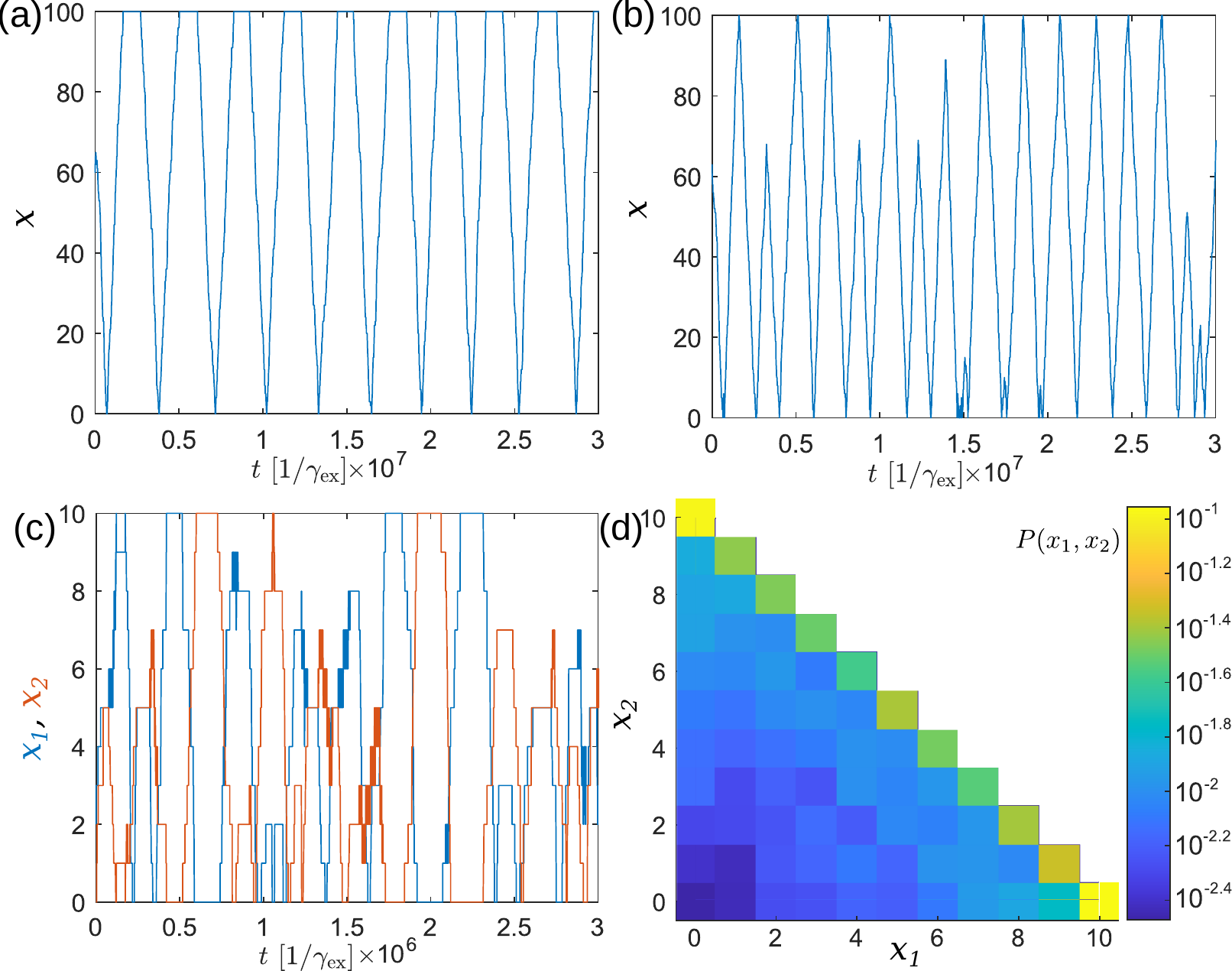}
 \caption{(a) Simulated stochastic trajectory for the symmetric 3-state model with $\gin=10^{-4}\gex$ and size $N_x=100$. We observe ``waiting times'' between growth and shrinkage. (b) When the upwards internal transition rate is increased to $\gamma_\mathrm{in}^\mathrm{BA}=10^{-2}\gex$, making the system asymmetric, waiting times between growth and shrinkage become negligible, and it is more likely for the system to stochastically unbind from the edge state during growth, switching to shrinkage. (c) Trajectories for two systems, each described by the 3-state model, coupled through the constraint $x_1+x_2 \leq N_x$ describing competition for the same pool of monomers. Here, both systems have identical, symmetric transition rates $\gin = 10^{-4} \gex$ and size $N_x=10$. (d) Probability distribution of finding a given $x_1$ and $x_2$ simultaneously, obtained from the same simulation, using a total of $10^7$ steps starting from a random initial state. We find that symmetric systems clearly show anti-phase synchronization, see also Movie 3 in the Supplemental Material \cite{suppmat}. \label{fig4}}
 \end{figure}

An interesting consequence of having asymmetric transition rates is that they affect the shape of the system oscillations over time. In particular, the typical timescale for moving along the edges is governed by the slower internal transition rates $\gin$, which constitute the bottleneck. As an example, in Fig.~\ref{fig4}(a,b) we show how oscillations in $x$ change in the 3-state model when we increase the rate for the upwards internal transition $\gamma_\mathrm{in}^\mathrm{BA}$ such that $\gamma_\mathrm{in}^\mathrm{BA} \gg \gamma_\mathrm{in}^\mathrm{AC}=\gamma_\mathrm{in}^\mathrm{CB}$ while keeping $\gex \gg \gamma_\mathrm{in}^\mathrm{BA}$. The apparent ``waiting times'' for which the number of subunits $x$ remains constant (vertical edge) are strongly reduced, and we obtain a system for which growth appears to be immediately followed by shrinkage. Moreover, the enhanced upwards internal transition leads to more frequent unbinding from the bottom edge, resulting in a stochastic switching to shrinkage before the right corner $x=100$ has been reached. When this 3-state model describes addition of GTP-bound monomers, conversion to GDP-bound monomers, and removal of GDP-bound monomers, this behavior is reminiscent of the dynamic instability of microtubules. Notably, our model is capable of demonstrating phases of growth followed by sporadic phases of shrinkage \cite{Mitchison1984,Leibler1993} using just three main timescales.

 Previous work has suggested that topological phonon modes may play a role in the dynamic instability of microtubules, describing vibrational modes that concentrate near the microtubule cap \cite{PhysRevLett.103.248101}. The topological states described therein relate to the elastic properties of a microtubule, and are independent of the polymerization or depolymerization dynamics. We propose, on the other hand, topological edge currents in the state space of a Markov model for the dynamics of microtubule polymerization and depolymerization. Hence, our models constitute new candidates for direct comparison with experimental data of microtubule lengths during catastrophe. 
 
 Intriguingly, the internal states (A, B, C) in our model could be related to distinct topological states in the elastic properties of the microtubule. For example, our state C, in which the microtubule is primed for depolymerization, could correspond to the topological state described previously \cite{PhysRevLett.103.248101} for which vibrational edge modes tend to open the microtubule end cap and induce depolymerization, whereas states A and B would correspond to states without protected vibrational edge modes, in which case the end cap remains intact and polymerization proceeds.
 

 \subsection{Coupled systems and synchronization \label{sec:coupled}}

In contrast with quantum topological systems, in which the boundaries represent real-space edges of a two-dimensional material and are thus fixed, the boundaries in stochastic systems represent constraints in configuration space, for example determined by the availability of subunits of a certain type in solution. This implies that the boundaries can dynamically change in time. In particular, if we have two systems (1 and 2), which are determined by their states $(x_1,y_1)$ and $(x_2,y_2)$, a global constraint on the number of subunits of type X would result in the constraint $x_1 + x_2 \leq N_x$. The boundaries for one system then depend on the state of the other system, i.e.~we have $0 \leq x_1 \leq N_x-x_2$ for system 1 and $0 \leq x_2 \leq N_x - x_1$ for system 2.

Remarkably, this boundary coupling can lead to synchronization (or entrainment) between the two systems. Stochastic simulations for the symmetric 3-state model, with constraints $y_1 \leq x_1$ and $y_2 \leq x_2$ for the second coordinate, show anti-phase synchronization between the $x$-coordinates of the two systems, see Fig.~\ref{fig4}(c,d). This coupled behavior emerges simply from shared physical constraints between the systems, without the need for direct interaction between them. Situations in which two or more systems compete for a limited pool of constituents arise naturally in the biological context e.g.~when multiple protein filaments are simultaneously polymerizing. Competition for a limited number of KaiA molecules is also hypothesized to underlie the synchronization of multiple KaiABC oscillators, through the phenomenon of differential affinity \cite{vanZon7420}.

Our results serve as a first demonstration of synchronization due to shared boundaries in topological systems. Nevertheless, much remains to be explored about the nature and mechanisms of synchronization in these systems. We anticipate that, by further studying the parameter space of these models, including asymmetric and weakly chiral systems, different kinds of synchronization (e.g.~in-phase synchronization) may be found. As a preliminary result, we find weak in-phase synchronization for an asymmetric system (see Appendix~\ref{app:synchro} and Fig.~\ref{extfig:synchro} there). Coupling of $n>2$ subsystems (through a constraint $\sum_i^n x_i \leq N_x$) is straightforward, and it will be interesting to explore the behavior of such systems in the large $n$ limit. More complex constraints may also be imposed, such as nonlinear constraints or constraints that mix the $x$ and $y$ coordinates of different subsystems. While here we have only considered coupling between two identical subsystems, studies of coupled non-identical systems with different transition rates will shed light on whether shared boundaries are capable of synchronizing not just the phases but also the oscillation frequencies of the different subsystems.

\section{Comparison with previous proposals \label{sec:comparison}}

We now put our results in the context of recent developments regarding topologically-protected states in related systems, namely classical systems with discrete states on a lattice \cite{muru17,DasbiswasE9031,2020arXiv200901780K,yoshida2020chiral}. These pertain to two main classes: (i) stochastic systems with linear dynamics described by a master equation, which includes Refs.~\citenum{muru17,DasbiswasE9031} as well as the present work, and (ii) nonlinear dynamical systems of the Lotka-Volterra type, which includes Refs.~\citenum{2020arXiv200901780K,yoshida2020chiral}.

With regards to stochastic systems, previous proposals \cite{muru17,DasbiswasE9031} only demonstrated topological states in one-dimensional lattices. In particular, these states showed static localization of probability at the system's edge, or at the boundary between two systems with distinct topological properties. Due to their low dimensionality, these systems obviously cannot support edge currents. Our work is thus the first demonstration of the emergence of topologically-protected chiral edge currents in such systems, which are of the highest relevance to biomolecular and biochemical processes, given that these are most often described by a stochastic master equation formalism. 

On the theoretical front, these previous proposals \cite{muru17,DasbiswasE9031} mapped the stochastic transition rate matrix (or more precisely, the part of it that corresponds to the generator of the probability flux along the periodic direction of the system, after separating it from the divergence operator) onto a corresponding Hermitian Hamiltonian, using a construction similar to that previously used for mechanical lattices \cite{Kane2014}. These mappings can change the resulting lattice structure, e.g.~Ref.~\citenum{DasbiswasE9031} shows a resulting Hamiltonian with higher-order couplings between next nearest neighbors, which are absent in the original transition matrix. The physical relation between the Hamiltonian and transition matrix, or indeed, any general connection they might have, is not entirely transparent. Instead, our work offers a general mapping of any transition matrix into a corresponding non-Hermitian Hamiltonian using the similarities between the master equation and the Schroedinger equation (see Appendix \ref{app:berry}). Our identification of the transition matrix without its diagonal part (which we show to be irrelevant in a periodic setting) as a typical electronic model allows access to the wealth of previous work in condensed matter theory. Here, our transition matrix retains the same symmetries and connections between neighbors as the electronic model, providing a clear physical relation between them, and a pathway to further progress. 

Even more recently, concepts of topological protection have been applied to nonlinear dynamical systems of the Lotka-Volterra type \cite{2020arXiv200901780K,yoshida2020chiral}, representing mass-conserving population dynamics. In these systems, multiple three-state loops representing predator-prey interactions of the rock-paper-scissors type are linked together to form a lattice. Similarly to the case of stochastic systems, for one-dimensional lattices \cite{2020arXiv200901780K} edge localization has been reported. On the other hand, chiral edge modes have been shown when the rock-paper-scissors loops are assembled into a two-dimensional Kagome lattice \cite{yoshida2020chiral}.

We note, however, some key differences (beyond the fundamental differences between a linear-stochastic and a nonlinear-deterministic system) between the chiral edge modes reported in Ref.~\citenum{yoshida2020chiral} and the chiral edge currents reported here. First, the model described in Ref.~\citenum{yoshida2020chiral} shows a uniform stationary state, and edge modes are only apparent in the dynamics of perturbations about this stationary state. Our model, on the other hand, can display arbitrarily strong localization at the edges of the system, even in stationary state. A second subtle, but significant difference regards the analogy to the quantum Hall effect. In Ref.~\citenum{yoshida2020chiral}, it is claimed that the individual rock-paper-scissors cycles behave like cyclotron orbits and give rise to the edge modes. However, one may note that, in their system, the chirality of the rock-paper-scissors cycles matches the chirality of the edge modes, so that e.g.~clockwise cycles result in clockwise edge modes. This is in contrast to the actual quantum Hall effect, in which clockwise cyclotron orbits give rise to counter-clockwise skipping orbits at the edges, see Fig.~\ref{fig1}(a). This behavior is preserved in our system, in which the chirality of external cycles (analogous to cyclotron orbits) is opposite to the chirality of the edge currents.

Interestingly, both our 4-state and 3-state lattices [see Fig.~\ref{fig1}(c,e)], including the different strengths of the internal and external transition rates $\gin$ and $\gex$, may be implemented within the same anti-symmetric Lotka-Volterra formalism used in Ref.~\citenum{2020arXiv200901780K}. Future work may thus explore how the edge currents we find here are realized in contexts relevant to population dynamics and game theory.

 \section{Conclusion and outlook}

In summary, we show how stochastic systems with out-of-equilibrium cycles at the microscopic scale support chiral edge currents and thus global cycles at the macroscopic scale, along the boundaries of the system's configuration space. The emergence of edge currents can be understood as a topological transition showing unique non-Hermitian properties, which highlight the nonequilibrium character of the system. The flexibility of the underlying configuration space affords exotic features not typically seen in topological systems, such as asymmetric transitions leading to dynamical instabilities, or coupling between subsystems through shared boundaries, leading to synchronization.

The models we propose here are not only interesting due to their application to biochemical systems, but also introduce novel topological phases. In this regard, we note that they are qualitatively different from previous extensions of the original 1d Hermitian SSH model.  For instance, our model contains propagating edge currents, whereas other extensions such as the 2d Hermitian \cite{PhysRevB.100.075437}, 1d non-Hermitian \cite{doi:10.1021/acsphotonics.8b00117,PhysRevLett.116.133903,PhysRevB.97.045106} or 2d non-Hermitian where non-Hermiticity comes directly from complex terms \cite{PhysRevA.100.032102}, only contain edge localization. We note that this is also the case with regards to the recent attempts to identify correspondence between stochastic systems and topological phases: such 1d models describe stationary localization without global currents \cite{muru17,DasbiswasE9031}. 

Our models exhibit rich phenomenology to be compared with various biochemical processes and other stochastic systems. In contrast to usual models for oscillatory processes at the biomolecular scale, which hard-code the desired global cycle into the structure of the reaction network, our models are constructed from identical repetitive motifs representing simple reactions at the level of single constituents. The global cycle arises as an emergent property of the system, and can occur over widely varying time scales and length scales simply by changing the number of constituents in the system, directly linking structure to emergent function.

In future work, it will be important to investigate and catalogue different oscillatory biochemical processes \cite{Li2018}, to elucidate whether topologically-protected edge currents have already been put to use in biological systems. This may prove a difficult task, however, as in practice a system deep in the topological phase ($\gex/\gin \ll 1$) behaves very similarly to a unicyclic oscillator that only runs along the edge states. A promising sign that a system may be exploiting topologically-protected edge currents will be the observation of occasional excursions into bulk states, and in particular the observation during these excursions of microscopic ``futile'' cycles that leave the system unchanged (e.g.~repeated single polymerization-depolymerization or phosphorylation-dephosphorylation steps). In the example of the KaiABC system, by excursions into the bulk we mean the observation of states in which any intermediate number of monomers (i.e.~between 1 and 5) have undergone a conformational change and been phosphorylated, see Fig.~\ref{fig:oscillations}(a). For our fully chiral model of the KaiABC system, our results from Sec.~\ref{sec:edgecurrents} imply that excursions into the bulk will be observed every $\approx \frac{1}{24}\frac{\gex}{\gin}$ cycles on average, which highlights again how observation of such excursions becomes increasingly difficult when the system is deeper in the topological phase. In this respect, it would be particularly useful if the ratio $\gex/\gin$ could be adjusted experimentally within a system of interest, which would be possible e.g.~if the external and internal transitions have distinct dependencies on a control parameter (such as concentration of ATP or GTP, pH, salt, or specific molecules known to affect function such as microtubule-targeting drugs \cite{Kaul2019}). Observation of a system that transitions from incoherent oscillations with frequent excursions into the bulk, to coherent oscillations with rare or no visits to the bulk, as a function of such a control parameter would constitute a strong indication of topologically-protected edge currents at play.

Lastly, our work opens up new directions for the theoretical exploration of new topological phases with interesting properties. For instance, our introduction of biologically-inspired features such as dynamical boundaries, new geometries and coupled systems suggest avenues for the realization of novel states. The consideration of heterogeneous transition rates or systems with disorder likewise merits further investigation. More broadly, the relation between general non-Hermitian features like exceptional points and vorticity, and their physical consequences in various systems, forms an emerging area of research. After all, our models suggest new topological states that can implemented in other, more easily tunable systems such as photonics \cite{PhysRevX.4.021017,Benalcazar61,doi:10.1021/acsphotonics.8b00117,PhysRevA.100.032102} and topoelectronics \cite{Imhof2018,PhysRevLett.122.247702}. By producing a blueprint for the emergence of protected edge currents, our formalism can be used to guide the design of new pathways in synthetic biology and materials self-assembly, e.g.~towards desired oscillatory assembly or reaction processes.  These rapid and continuing developments hold promise for the prediction of new states of matter in both classical and quantum systems. 

\acknowledgements

We thank James Sethna and Beno\^{i}t Mahault for helpful discussions. Talks and interactions during the KITP Program \textit{Symmetry, Thermodynamics and Topology in Active Matter} were also stimulating and thought-provoking.
This study was supported by the Max Planck Society. \\

\appendix

\section{Stationary state in the fully chiral case \label{app:stationary1}}

We first directly analyze the steady state of the system in the fully chiral case, with $\gpin=\gpex=0$. We take the bottom edge of the lattice, be it square or Kagome, as an example without loss of generality. We call the probability for a B site on the edge $p_B$, and for the C site $p_C$. The probability of bulk sites away from the edge is called $p_b$. The stationarity condition for the B site reads $\gin p_C - \gin p_B - \gex p_B = 0$ whereas for the C site it reads $\gin p_b + \gex p_B - \gin p_C = 0$. Using both to solve for $p_B$ and $p_C$, we obtain $p_B = p_b$ and $p_C = \frac{\gin + \gex}{\gin} p_b$. The fact that $p_B = p_b$ ensures that the bulk site contiguous to the edge site B is stationary as well, with probability $p_b$, as are all other bulk sites.

 \begin{figure}
 \includegraphics[width=0.7\linewidth]{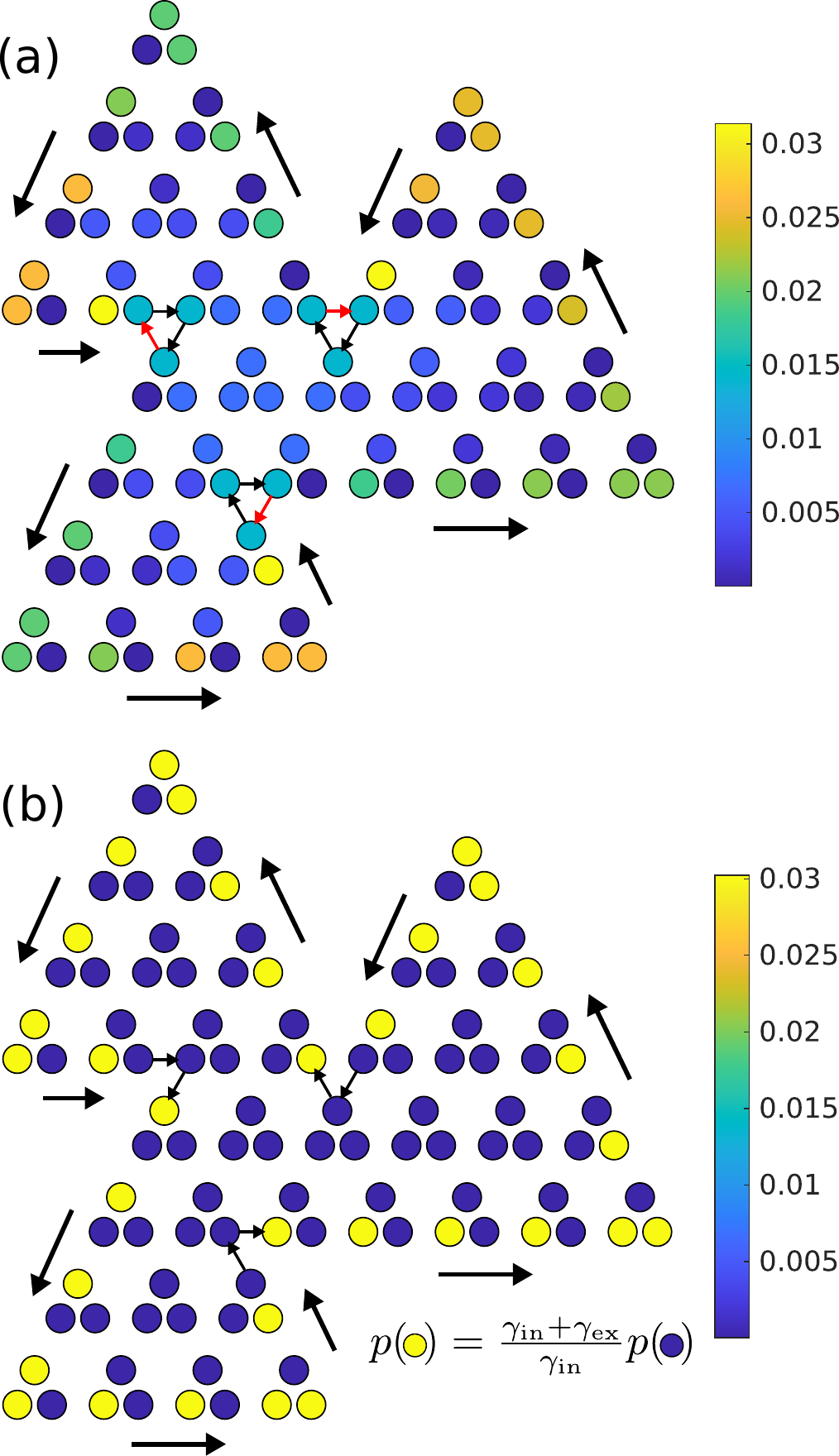}
 \caption{Edge currents and the resulting cycles are robust with respect to the shape of the boundaries, also in the case of the 3-state (Kagome) lattice. However, in contrast to the 4-state model, in this case one must make sure that corner-cutting transitions across concave corners of the lattice are absent, as otherwise such corners scatter probability into the bulk. In (a), corner-cutting transitions are present (red arrows), and the three concave corners scatter probability into the bulk, making the steady-state probabilities sensitive to the shape of the system. In (b), corner-cutting transitions are absent, the edge current does not scatter, and the steady-state probabilities are independent of the shape of the system. The steady states have been calculated from direct solution of the master equation, for a fully-chiral system with $\gin=10^{-3}\gex$.    \label{extfig:kagome_problem}}
 \end{figure}

Stationarity further implies that sites on convex corners (which have an incoming and an outgoing internal transition) have probability $p_C$ as well, see Fig.~\ref{fig2}(d) and Fig.~\ref{extfig:kagomesim}(d). On the other hand, sites on concave corners (which have an incoming and an outgoing external transition, and appear on non-rectangular 4-state and non-triangular 3-state geometries) have probability $p_b$, see Fig.~\ref{fig2}(e) and Fig.~\ref{extfig:kagome_problem}(b). Note, however, that the latter is true for the 3-state (Kagome) model only in the absence of corner-cutting transitions, see Fig.~\ref{extfig:kagome_problem}. If such transitions are present, the edge probability scatters into the bulk, the stationarity condition can no longer be determined using local balance only, and the stationary probability distribution becomes dependent on the system shape and size (and no longer given by just a combination of two values $p_b$ and $p_C = \frac{\gin + \gex}{\gin} p_b$).

The probability current along the edge can be calculated as $J_\mathrm{edge} = \gin p_C - \gin p_b = \gex p_b$. To obtain the global probability of being at the edge in a square 4-state model with $N_x=N_y=N$, we note that there are $n_{e,C}=4(N+1)$ sites with probability $p_C$ on the edge, $n_{e,b}=4N$ sites with probability $p_b$ on the edge, and $n_b=4N^2$ bulk sites, all with probability $p_b$. Normalizing the total probability in the system by setting $p_C n_{e,C} + p_b (n_{e,b} + n_b) = 1$, we can calculate $p_b$ as
\begin{equation}
    p_b = \frac{1}{4(N+1)} \frac{1}{N+\frac{\gin+\gex}{\gin}}
\end{equation}
The global probability of being at the edge is then calculated as $P_\mathrm{edge}=p_C n_{e,C} + p_b n_{e,b}=1-p_b n_b$ or, explicitly,
\begin{equation}
    P_\mathrm{edge} = \frac{ \frac{\gin + \gex}{\gin} (N+1) + N }{ \left( N + \frac{\gin + \gex}{\gin} \right) (N+1) }
\end{equation}
which in the limit $\gex \gg \gin$ results in the expression quoted in Section \ref{sec:edgecurrents}.

To calculate the current $J_|$, i.e.~the steady-state current going through the dashed line in the inset of Fig.~\ref{fig:oscillations}, we note that it can be written as $J_| = \gin p_C  = (\gex + \gin) p_b = J_\mathrm{edge}+\gin p_b$, where the second term represents the contribution from the internal loop at the center of the lattice. Introducing the explicit expression for $p_b$, we obtain the result in Eq.~(\ref{eq:T}) in the main text.

\section{Symmetries and band structure of $\mc{W}$ \label{app:symmetries}}

 \begin{figure}
 \includegraphics[width=\linewidth]{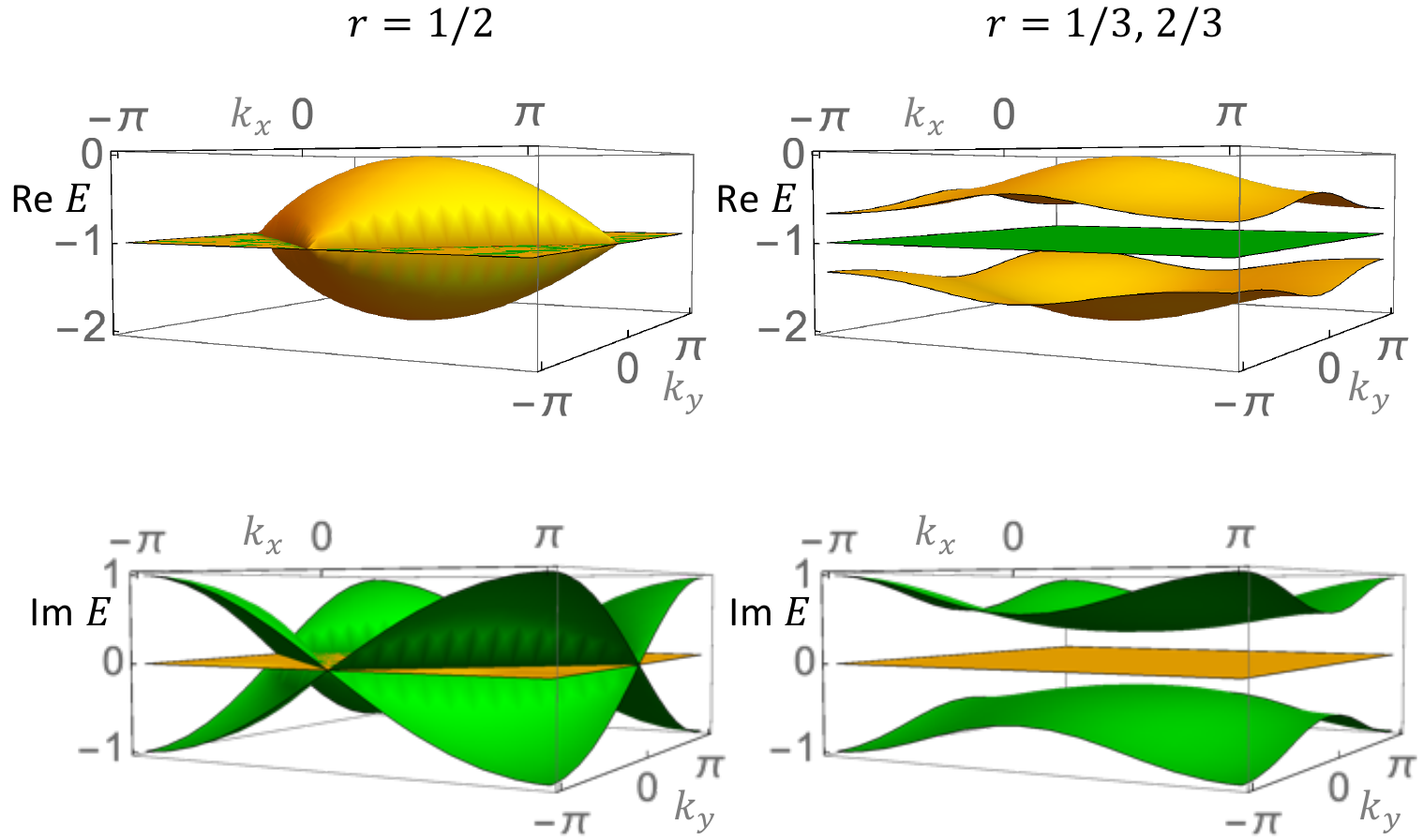}
 \caption{Band structure of $\mc{W}_{\vect{k}}$ in the fully chiral case. The 
 bands (or spectrum) $E(\vect{k})$ can be exactly solved for the periodic system (see Eq.~\ref{eq:periodicspectc}), where the top $E_{+,+}$ and bottom $E_{-,+}$ bands are in yellow and the middle bands $E_{+,-}$ and $E_{-,-}$ are in green. \textit{Left}: When $r=1/2$ ($\gamma_{\textrm{in}}=\gamma_{\textrm{ex}}$), the bands have degeneracies at $E(\vect{k})=-\gamma_\mathrm{tot}$. \textit{Right}: When  $r\neq 1/2$ ($\gamma_{\textrm{in}}\neq \gamma_{\textrm{ex}}$),  band gaps open in real space (top) and  imaginary space (bottom) away from $E=-\gamma_\mathrm{tot}$. As the spectrum is even about $\gamma_{\textrm{in}}=\gamma_{\textrm{ex}}$, the same result is obtained for $r=1/3$ and $2/3$, which are the parameters we use. Note that the ratio $r$ interpolates between the transition probability strengths, i.e.~$\gamma_{\textrm{ex}}=r\gamma_\mathrm{tot}$ and  $\gamma_{\textrm{in}}=(1-r)\gamma_\mathrm{tot}$. In all panels, values of $E$ are given in units of $\gamma_\mathrm{tot}$. \label{extfig:periodicbands}}
 \end{figure}
 The properties of the transition matrix $\mc{W}$ governing the Master equation can be analyzed using the decomposition $\mc{W}=\mc{A}-\mc{D}$, where $\mc{A}_{ij}=\langle i|j\rangle$ is the transition rate from state $p_j$ to $p_i$ and $\mc{D}_{ij}=\delta_{ij}\sum_k\langle k|i\rangle$ \cite{RevModPhys.48.571}. $\mc{A}$ and $\mc{W}$ are also the adjacency and Laplacian matrices, respectively. 
 
 For the 4-state model in Fig.~\ref{fig1}(c), we can write the adjacency matrix $\mc{A}$ explicitly, where we denote the four internal states of cell $(x,y)$ as $|(x,y)_A\rangle$ to $|(x,y)_D\rangle$, as
\begin{widetext}
\begin{eqnarray}
\mc{A}&=&\gamma_{\textrm{in}}\sum^{N_x}_{x=0}\sum^{N_y}_{y=0}\left[|(x,y)_B\rangle\langle(x,y)_C|+|(x,y)_A\rangle\langle(x,y)_B|+|(x,y)_D\rangle\langle(x,y)_A|+|(x,y)_C\rangle\langle(x,y)_D|\right]\nonumber\\
&+&\gamma_{\textrm{ex}}\{\sum^{N_x-1}_{x=0}\sum^{N_y}_{y=0}\left[|(x+1,y)_C\rangle\langle(x,y)_B|+|(x,y)_A\rangle\langle(x+1,y)_D|\right]+\sum^{N_x}_{x=0}\sum^{N_y-1}_{y=0}\left[|(x,y+1)_B\rangle\langle(x,y)_A|+|(x,y)_D\rangle\langle(x,y+1)_C|\right]\}\nonumber\\
&+&\gamma_{\textrm{ex}}^0\{\sum^{N_y}_{y=0}\left[| (0,y)_C\rangle\langle (N_x,y)_B|+|(N_x,y)_A\rangle\langle(0,y)_D|\right]+
\sum^{N_x}_{x=0}\left[|(x,0)_B\rangle\langle(x,N_y)_A|+|(x,N_y)_D\rangle\langle(x,0)_C|\right]\}.
\end{eqnarray}
\end{widetext}
The edge transition probability $\gamma_{\textrm{ex}}^0$ interpolates between periodic boundary conditions (PBC) and open boundary conditions (OBC). PBC occur when $\gamma_{\textrm{ex}}^0=\gamma_{\textrm{ex}}$, while OBC occur when $\gamma_{\textrm{ex}}^0=0$. The diagonal matrix $\mc{D}$ can then be easily calculated using its definition above.

This describes a 2d non-Hermitian version of the SSH model \cite{PhysRevLett.42.1698}, which we analyze here for simplicity (the more general case will follow). For a system with PBC, this transition matrix can be expressed in Fourier space as
\[
\mc{W}_{\vect{k}}=
  \begin{pmatrix}
  -\gamma_{\textrm{tot}}& \gamma_{\textrm{in}}& 0&\gamma_{\textrm{ex}}e^{-ik_x}\\
  \gamma_{\textrm{ex}}e^{ik_y}&  -\gamma_{\textrm{tot}}&\gamma_{\textrm{in}}&0\\
     0& \gamma_{\textrm{ex}}e^{ik_x}&  -\gamma_{\textrm{tot}}&\gamma_{\textrm{in}}\\
        \gamma_{\textrm{in}}& 0&\gamma_{\textrm{ex}}e^{-ik_y}&  -\gamma_{\textrm{tot}}
  \end{pmatrix}.
\]
where $\gamma_{\textrm{tot}}=\gex+\gin$ and $\vect{k}=(k_x,k_y)$, the reciprocal lattice vector.

$\mc{W}_{\vect{k}}$ obeys inversion and time-reversal symmetries,  $\mc{I}\mc{W}_{\vect{k}}\mc{I}^{-1}=\mc{W}_{\vect{-k}}$ and $\mc{W}_{\vect{k}}=\mc{W}_{-\vect{k}}^*$ respectively. $\mc{I}$ is a unitary operator which can be represented as $\mc{I}=\sigma_x\otimes\mathbb{1}$, where $\sigma_x$ is a Pauli matrix and $\mathbb{1}$ is the identity matrix. In addition, using $\mc{W}_{\vect{k}}=\mc{A}_{\vect{k}}-\gamma_{\textrm{tot}}\mathbb{1}$, we note that $\mc{A}_{\vect{k}}$ further obeys sublattice symmetry $\mc{S}\mc{A}_{\vect{k}}\mc{S}^{-1}=-\mc{A}_{\vect{k}}$, where $S$  can be represented as $\mc{S}=\mathbb{1}\otimes\sigma_z$.


With regards to the spectrum, these symmetries suggest that the eigenvalues are either real or complex with conjugate pairs. We indeed see this when analyzing the spectrum of $\mc{W}_{\vect{k}}$:
\begin{eqnarray}
E(\vect{k})_{\pm,\pm}= -\gamma_{\textrm{tot}}\pm\sqrt{a(\vect{k})\pm\sqrt{a(\vect{k})^2+\beta^2}}\label{eq:periodicspectc}
\end{eqnarray}
where $a(\vect{k})=\gamma_{\textrm{in}}\gamma_{\textrm{ex}}(\cos k_x+\cos k_y)$ and $\beta=(\gamma_{\textrm{in}}^2-\gamma_{\textrm{ex}}^2)$. 
Each pair is illustrated in yellow and green respectively in Fig.~\ref{extfig:periodicbands}. The spectrum is even about  $\gamma_{\textrm{in}}=\gamma_{\textrm{ex}}$, where bandgaps open for $\gamma_{\textrm{in}}\neq \gamma_{\textrm{ex}}$. At $\gamma_{\textrm{in}}=\gamma_{\textrm{ex}}$, the bandgap closes to yield degenerate solutions at $E(\vect{k})=-\gamma_{\textrm{tot}}$. 

We can similarly obtain the spectrum for the 3-state system [Fig.~\ref{fig1}(e)] which obeys the expression
\begin{eqnarray}
&&\left(E(\vect{k})+\gamma_{\textrm{tot}}\right)^3+ \gamma_{\textrm{in}}^3+\gamma_{\textrm{ex}}^3=\nonumber\\& & \gamma_{\textrm{in}}\gamma_{\textrm{ex}}\left[e^{ik_x}+2e^{-i\frac{k_x}{2}}\cos \left(\frac{\sqrt{3}k_y}{2}\right)\right](E(\vect{k})+\gamma_{\textrm{tot}}).\nonumber
\end{eqnarray}

Upon generalizing the phase space to include the reverse transitions, the spectrum of $\mc{W}_{\vect{k}}$ still holds a similar form as Eq.~\eqref{eq:periodicspectc} and is given in the main text. The symmetries of $\mc{W}_{\vect{k}}$ and $\mc{A}_{\vect{k}}$ that were previously discussed also remain in the general case. 

\section{Berry connection and Zak phase of $\mc{W}$ \label{app:berry}}

The Berry connection is defined for the Hamiltonian $\mc{H}$ as the generator of time translation, i.e.
\begin{eqnarray}
-i\frac{d}{dt}\psi(t)=\mc{H}\psi(t).\label{eq:schro}
\end{eqnarray}

In this section, we show that the Berry connection and Zak phase  can be identically computed using the eigenvectors of $\mc{W}$. Consider the Berry connection  $Q_m(\vect{k})$ defined for a Hermitian Hamiltonian \cite{PhysRevLett.62.2747,PhysRevB.100.075437} as
\begin{eqnarray}
Q_m(\vect{k})=i\psi^{\dagger}_m(\vect{k})\partial_{\vect{k}}\psi_m(\vect{k}).
\end{eqnarray}
Here the $\psi_m$s are the right eigenvectors of the Hamiltonian. For a non-Hermitian Hamiltonian, this becomes a complex quantity \cite{PhysRevB.97.045106,doi:10.1021/acsphotonics.8b00117}
\begin{eqnarray}
Q^c_m(\vect{k})=i\phi^{\dagger}_m(\vect{k})\partial_{\vect{k}}\psi_m(\vect{k}).\label{eq:berry}
\end{eqnarray}
where now $\psi$ and $\phi$ are normalized biorthogonal right and left eigenvectors. The latter are also the eigenvectors of the Hermitian conjugate of the Hamiltonian.

Now, since the Master equation is identical to the Schroedinger equation in Eq.~\eqref{eq:schro} up to a prefactor, an effective Hamiltonian $\mc{H}$ can be defined which is identical to $\mc{W}$ up to a prefactor. $\mc{W}$ and $\mc{H}$ have the same eigenvectors, and the Berry connection defined above will be identical to the Berry connection of $\mc{W}$. 

Using the decomposition mentioned above of $\mc{W}=\mc{A}-\mc{D}$ allows further progress. Under periodic boundary conditions, $\mc{D}$ is simply proportional to the identity, so its presence does not change the resulting eigenvectors. Thus the eigenvectors of $\mc{W}$ and $\mc{A}$ are also identical, as is the Berry connection of their respective bands. Further, $\mc{A}$ obeys sublattice (or chiral) symmetry $\mc{S}\mc{A}_{\vect{k}}\mc{S}^{-1}=-\mc{A}_{\vect{k}}$, which has been shown to quantize the integral of the complex Berry connection $Q^c_m(\vect{k})$ across reciprocal space \cite{PhysRevB.97.045106,doi:10.1021/acsphotonics.8b00117}. When the integral is 0, the system is in the trivial phase and will mostly remain in the bulk. When the integral is $\pi$, the system is in the topological Zak phase and edge states will dominate. Since $\mc{W}$ inherits this Berry connection quantization, it therefore supports the Zak phase and long-lived edge dynamics.

$\mc{W}$ also has inversion symmetry, which similarly quantizes the integrated Berry connection in the Hermitian limit \cite{PhysRevB.100.075437,PhysRevLett.62.2747}. This limit has been well-studied, where previous results show that $\mc{A}$ (the 2d SSH model  \cite{PhysRevB.100.075437}) and therefore $\mc{W}$ exhibit the Zak phase above $r^*=1/2$. We verify that the bandgap does not close as we interpolate from $c=1/2$ into the fully non-Hermitian limit $c=1$ above $r^*(c)$, using our expression for the band structure of $\mc{W}$ given in the main text. As the system maintains sublattice symmetry throughout and the bands remain separated above $r^*(c)$, it exhibits the Zak phase and edge states. 
 \begin{figure}
 \includegraphics[width=0.7\linewidth]{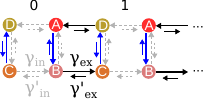}
 \caption{Schematic of half-periodic geometry. The system is periodic along the vertical direction, and open along the horizontal direction. The stationary probability for the four sites belonging to a cell a distance $n$ away from the left edge is given by the 4-vector $\boldsymbol{P}_n \equiv [p_D^{n}~p_C^{n}~p_A^{n}~p_B^{n}]^T$. The solid blue arrows denote the $\gex$ and $\gpex$ transitions that fold back along the periodic direction.   \label{extfig:transfermat}}
 \end{figure}

\section{Stationary state of $\mc{W}$ in the general case \label{app:stationary2}}

We now analyze the steady state of the general system, including the reverse transitions $\gpex$ and $\gpin$. We focus on the 4-state system, and consider a ribbon periodic along the vertical dimension but open along the horizontal direction (Fig.~\ref{extfig:transfermat}). Defining the vectors $\boldsymbol{P}_n^- \equiv [p_D^{n}~p_C^{n}]^T$ and $\boldsymbol{P}_n^+ \equiv [p_A^{n}~p_B^{n}]^T$, the stationarity conditions can be written as
\begin{eqnarray}
\boldsymbol{P}_n^- &=& U_1 \boldsymbol{P}_{n-1}^- + U_2 \boldsymbol{P}_{n-1}^+  \\
\boldsymbol{P}_n^+ &=& U_3 \boldsymbol{P}_{n-1}^+ + U_4 \boldsymbol{P}_n^- 
\end{eqnarray}
where we have defined the matrices
\begin{equation}
U_1 \equiv \begin{pmatrix}
-\gpin/\gex & 0\\
0 & -\gin/\gpex 
\end{pmatrix}
\end{equation}
\begin{equation}
U_2 \equiv \begin{pmatrix}
\frac{\gin + \gex + \gpin + \gpex}{\gex} & -\frac{\gin+\gpex}{\gex}\\
-\frac{\gex+\gpin}{\gpex} & \frac{\gin + \gex + \gpin + \gpex}{\gpex} 
\end{pmatrix}
\end{equation}
\begin{equation}
U_3 \equiv \begin{pmatrix}
-\gpex/\gin & 0\\
0 & -\gex/\gpin 
\end{pmatrix}
\end{equation}
\begin{equation}
U_4 \equiv \begin{pmatrix}
\frac{\gin + \gex + \gpin + \gpex}{\gin} & -\frac{\gpin+\gex}{\gin}\\
-\frac{\gpex+\gin}{\gpin} & \frac{\gin + \gex + \gpin + \gpex}{\gpin} 
\end{pmatrix}
\end{equation}

Plugging in the equation for $\boldsymbol{P}_n^-$ into the one for $\boldsymbol{P}_n^+$, we obtain a transfer matrix $M$ in the rightwards direction for the probability of the 4-site cells $\boldsymbol{P}_n \equiv [p_D^{n}~p_C^{n}~p_A^{n}~p_B^{n}]^T$, that is, the $4 \times 4$ matrix $M$ that gives
\begin{equation}
\boldsymbol{P}_n = M \boldsymbol{P}_{n-1}
\end{equation}
and has the form
\begin{equation}
M \equiv \begin{pmatrix}
U_1 & U_2\\
U_4 U_1 & U_4 U_2 + U_3 
\end{pmatrix}
\end{equation}

Two of the eigenvalues of $M$ are always equal to 1, and have identical associated eigenvectors $\boldsymbol{V}_1=[1~1~1~1]^T$. This reflects that the steady state is uniform in the bulk. However, we also find two other eigenvalues $\alpha$ and $1/\alpha$, with $0 \leq \alpha < 1$. The corresponding eigenvectors, $\boldsymbol{V}_\alpha$ and $\boldsymbol{V}_{1/\alpha}$, are non-trivial, and they are related to each other by
\begin{equation}
\boldsymbol{V}_{1/\alpha} = \begin{pmatrix}
0 & 0 & 0 & 1\\
0 & 0 & 1 & 0\\
0 & 1 & 0 & 0\\
1 & 0 & 0 & 0
\end{pmatrix} \boldsymbol{V}_{\alpha}
\end{equation}
i.e., $\boldsymbol{V}_\alpha$ is identical to $\boldsymbol{V}_{1/\alpha}$, except for a left-right, up-down reflection (parity symmetry). These properties strongly suggest that they correspond to the perturbations to the bulk behavior induced by the presence of the left edge ($\boldsymbol{V}_{\alpha}$) and right edge ($\boldsymbol{V}_{1/\alpha}$) of the system. The perturbation decays geometrically, with rate $\alpha$, as we move away from the edge.

   \begin{figure}
 \includegraphics[width=1\linewidth]{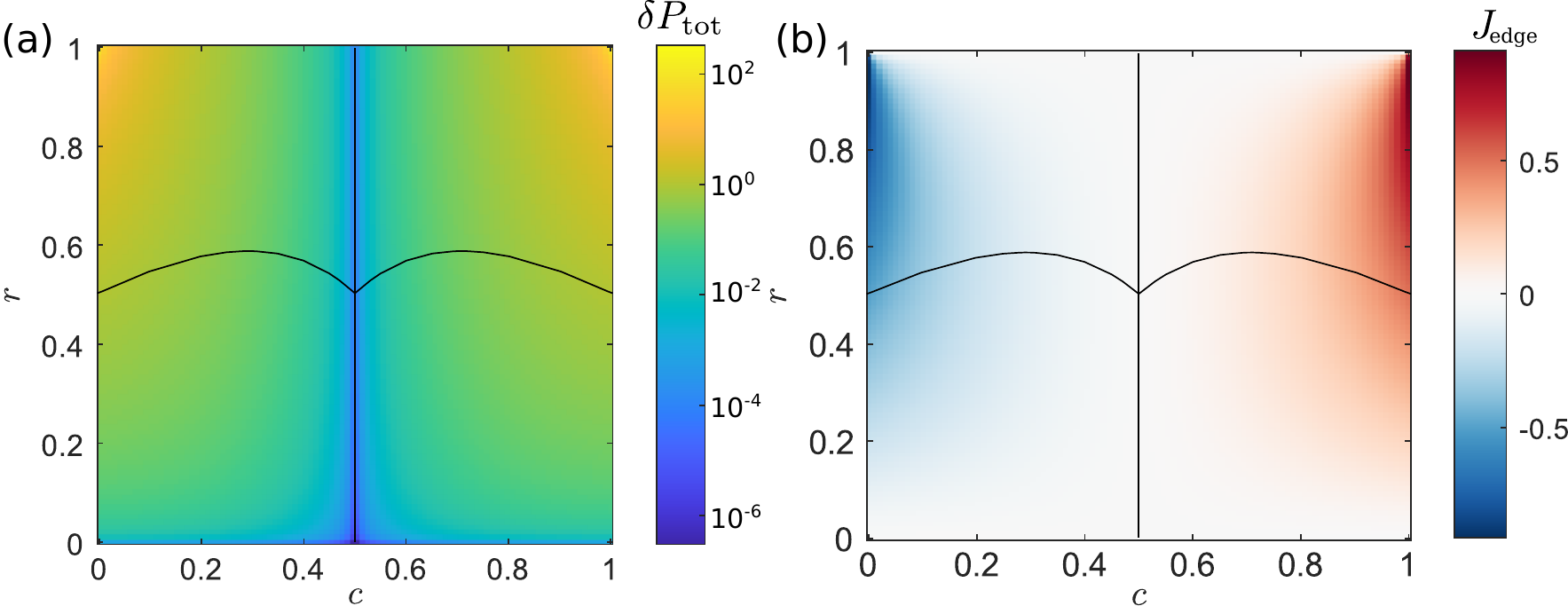}
 \caption{Results of the transfer matrix analysis of $\mc{W}$. (a) Probability disturbance $\delta P_\mathrm{tot}$ at the edge (in logarithmic scale) as a function of the chirality parameter $c$ and the ratio parameter $r$. (b) Edge flux $J_\mathrm{edge}$ as a function of the same parameters. Both the probability disturbance and the flux vanish in the achiral, Hermitian case $c=0.5$, as well as in the limit $r \to 0$. The flux also vanishes in the limit $r \to 1$. Otherwise, the probability is always accumulated at the edge ($\delta P_\mathrm{tot}>0$). The flux is positive (counter-clockwise) for $c>0.5$ and negative (clockwise) for $c<0.5$. Note that $\delta P_\mathrm{tot}$ is given in units of $p_b$, and $J_\mathrm{edge}$ in units of $p_b\gamma_\mathrm{tot}$. The black curves represent the boundary $r^*(c)$ separating the topological and trivial phases of the system. \label{extfig:fluxesphasediag}}
 \end{figure}
 
To ensure that these perturbations indeed correspond to stationary solutions at the edges, we try a solution of the form $\boldsymbol{P}_0 = p_b (  \boldsymbol{V}_1 + \xi \boldsymbol{V}_\alpha )$ at the left edge, where $p_b$ corresponds to the probability in the bulk, far away from the edge. We find that the stationarity conditions at the left edge are satisfied if
\begin{equation}
\xi = \frac{\gex - \gpex}{[(\gin+\gpin+\gpex)~-(\gpin+\gex)~-\gin~~0]\cdot \boldsymbol{V}_\alpha} \label{xi}
\end{equation}
The excess (or lack) of probability at the edge is therefore $\delta \boldsymbol{P}_0 = p_b \xi \boldsymbol{V}_\alpha$, while for the $n$-th cell away from the edge it is $\delta \boldsymbol{P}_n = p_b \xi \boldsymbol{V}_\alpha \alpha^n$. Assuming that the system size $N$ is large enough such that the probability disturbance decays away from the boundary, i.e.~$\alpha^N \ll 1$ or $N\gg -1/\log \alpha$, the total probability disturbance due to the presence of the edge can be calculated as
\begin{equation}
\delta P_\mathrm{tot} = \sum_{n=0}^\infty [1~1~1~1] \cdot  \delta \boldsymbol{P}_n = p_b \frac{\xi}{1-\alpha} [1~1~1~1] \cdot \boldsymbol{V}_\alpha
\end{equation}
which is positive if probability accumulates at the boundary, and negative if probability is depleted at the boundary. The total probability flux along the edge can be directly calculated from the steady state probabilities as
\begin{equation}
J_\mathrm{edge} = p_b \frac{\xi}{1-\alpha}~ [\gin~-\gpin~~\gpin~-\gin] \cdot \boldsymbol{V}_\alpha
\end{equation}
and is positive for net counter-clockwise edge flux (net flux downwards at the left edge) and negative for net clockwise edge flux (net flux upwards at the left edge).

 \begin{figure}
 \includegraphics[width=1\linewidth]{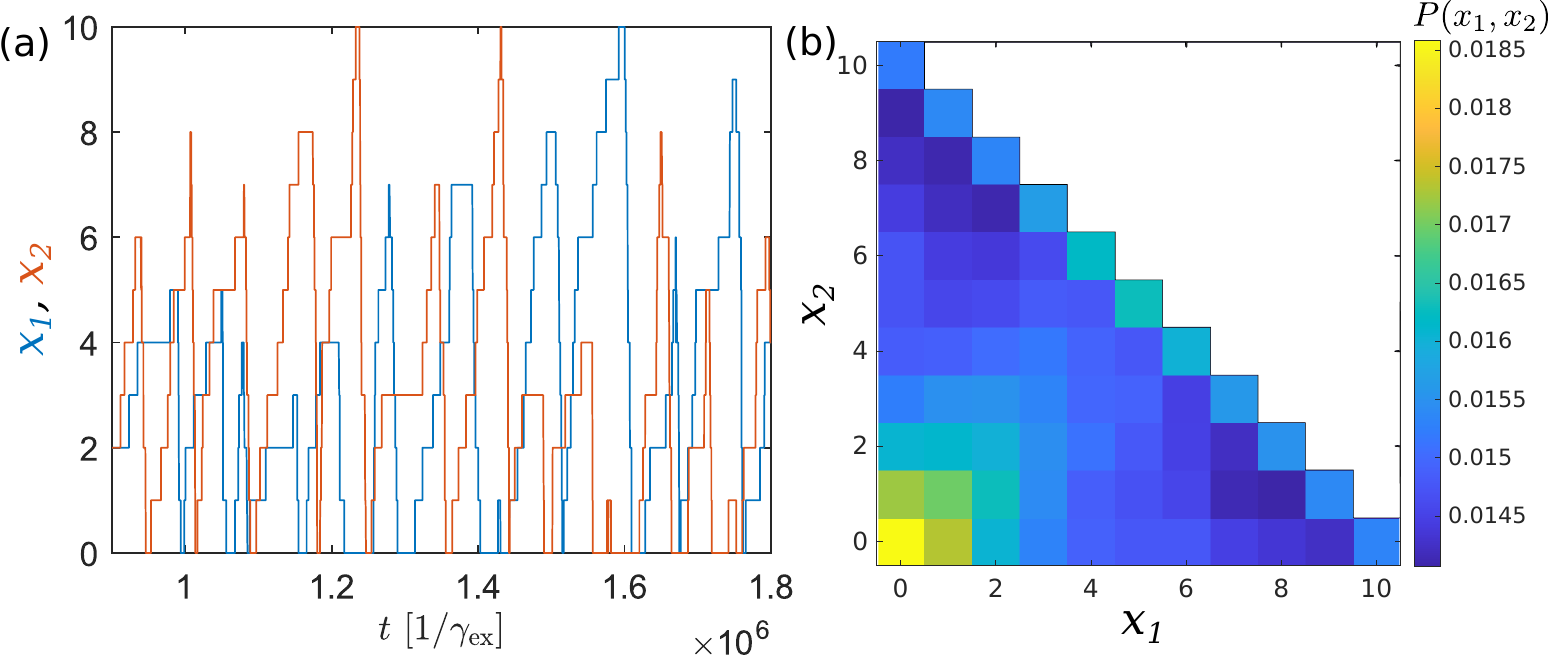}
 \caption{In-phase synchronization in an asymmetric 3-state model. (a) Trajectories for two coupled 3-state models, both with identical, asymmetric internal transition rates $\gin$ given by $\gamma_\mathrm{in}^\mathrm{CB}=10^{-4} \gex$ for the horizontal transitions, $\gamma_\mathrm{in}^\mathrm{AC}=10^{-3} \gex$ for the diagonal transitions, and $\gamma_\mathrm{in}^\mathrm{BA}=10^{-2} \gex$ for the upwards transitions. (b) Probability distribution of finding a given $x_1$ and $x_2$ simultaneously for the same simulation, using a total of $10^7$ steps starting from a random initial state. We find that the two systems show weak in-phase synchronization, particularly at initial growth. The system size is $N_x=10$. \label{extfig:synchro}}
 \end{figure}

In the limit of a fully chiral system, we find $\alpha=0$ and we recover the results obtained in Appendix~\ref{app:stationary1}. It is also interesting to note that, according to Eq.~(\ref{xi}), the effect of the boundaries completely vanishes (both in terms of probability disturbance and probability flux) when $\gex = \gpex$. Thus, chirality in the external transitions is essential to obtain boundary effects at steady state.

In this way, we can characterize the steady states of the system by simply studying the eigenvalues and eigenvectors of a $4 \times 4$ matrix. Notably, the results are independent of the system size or the shape of the boundaries, even if they give us information about probability accumulation and fluxes at the edges, see Fig.~\ref{extfig:fluxesphasediag}.

\section{Synchronization in systems with asymmetric rates \label{app:synchro}}

Due to the high dimensionality of the parameter space in asymmetric systems (8 transition rates in the fully chiral 4-state model and 6 transition rates in the fully chiral 3-state model, which double when we consider weakly chiral models), a systematic exploration of the possible types of synchronization in these models is a difficult task which we do not attempt in this work. As a proof of concept, however, we have simulated an asymmetric 3-state model, with symmetric external transitions but with internal transition rates fastest along the vertical direction, slower along the diagonal direction, and slowest along the horizontal direction ($\gamma_\mathrm{in}^\mathrm{BA} > \gamma_\mathrm{in}^\mathrm{AC} > \gamma_\mathrm{in}^\mathrm{CB}$), see Fig.~\ref{extfig:synchro}. We find weak in-phase synchronization, in particular during the initial stages of growth, i.e.~at low $x$.

 \bibliography{topo_chem}

\setcounter{figure}{0}
\renewcommand{\thefigure}{S\arabic{figure}}

\clearpage

\begin{widetext}

  \section*{Supplemental Videos}

\begin{itemize}
    \item Movie 1: Stochastic simulation of the fully-chiral, symmetric 4-state model with $\gex=10^3 \gin$ and $N_x=N_y=20$, for a total time $t\gin=550$. The movie shows a full lattice representation, an $(x,y)$-space representation, and a polymer-style representation, where blue and red monomers are counted by $x$ and $y$, respectively.
    \item Movie 2: Stochastic simulation of the fully-chiral, symmetric 3-state model with $\gex=10^3 \gin$, $N_x=20$ and the phosphorylation-type constraint $y\leq x$, for a total time $t\gin=550$. The movie shows a full lattice representation, an $(x,y)$-space representation, and a polymer-style representation, where the total number of monomers is counted by $x$, and the number of modified monomers (blue to red) is counted by $y$.
    \item Movie 3: Stochastic simulations of two coupled fully-chiral, symmetric 3-state models with $\gex=10^3 \gin$ and constraints $x_1 + x_2 \leq 20$, $y_1 \leq x_1$, $y_2 \leq x_2$,  for a total time $t\gin=550$. The movie shows a full lattice representation of each system, a combined $(x,y)$-space representation for both systems, and a polymer-style representation for each system, where the total number of monomers is counted by $x$, and the number of modified monomers (blue to red) is counted by $y$. In the full lattice representation, the dynamically changing boundaries for each system as a function of the state of the other system are represented by the grey shading (inaccessible region), with the accessible region being $x_1 \leq 20 - x_2$ for system 1, and $x_2 \leq 20-x_1$ for system 2. In the $(x,y)$-space representation, the blue dot corresponds to $(x_1,y_1)$ whereas the green dot corresponds to $(20-x_2,y_2)$. The blue dot thus moves counter-clockwise whereas the green dot moves clockwise, and the constraint $x_1+x_2 \leq 20$ implies that the blue dot must remain to the left of the green dot.
\end{itemize}

 \end{widetext}

\end{document}